\documentclass[twocolumn,preprintnumbers,amsmath,amssymb,superscriptaddress,nofootinbib,longbibliography]{revtex4-1}

\usepackage{graphicx}
\usepackage{bm}
\usepackage{dsfont}
\usepackage[usenames,dvipsnames]{xcolor}
\usepackage{pstricks}
\usepackage[tight]{subfigure}
\usepackage{verbatim}
\usepackage{units}
\usepackage{multirow}
\usepackage{enumitem}
\usepackage{mathrsfs}
\usepackage{leftidx}
\usepackage{xspace}
\usepackage{braket}
\usepackage{bbm}
\usepackage{amsfonts,amssymb,amsmath}

\usepackage[a4paper]{hyperref}
\hypersetup{colorlinks=true,linktoc=all,linkcolor=blue,breaklinks=true,citecolor=blue,urlcolor=blue}

\usepackage[english]{babel}
\usepackage[normalem]{ulem}

\usepackage{hyperref}

\AtBeginDocument{%
    \newwrite\bibnotes
    \def\bibnotesext{Notes.bib}
    \immediate\openout\bibnotes=\jobname\bibnotesext
    \immediate\write\bibnotes{@CONTROL{REVTEX41Control}}
    \immediate\write\bibnotes{@CONTROL{%
    apsrev41Control,author="08",editor="1",pages="1",title="0",year="1"}}
     \if@filesw
     \immediate\write\@auxout{\string\citation{apsrev41Control}}%
    \fi
}%

\begin{document}
	
	
	\title{Phase-coherent heat circulators with normal- or superconducting contacts}
	
	\author{Matteo Acciai}
	\affiliation{Department of Microtechnology and Nanoscience (MC2), Chalmers University of Technology, S-412 96 G\"oteborg, Sweden}

	\author{Fatemeh Hajiloo}
	\affiliation{Department of Microtechnology and Nanoscience (MC2), Chalmers University of Technology, S-412 96 G\"oteborg, Sweden}
	
	\author{Fabian Hassler}
	\affiliation{JARA Institute for Quantum Information, RWTH Aachen University, 52056 Aachen, Germany}
	
	\author{Janine Splettstoesser}
	\affiliation{Department of Microtechnology and Nanoscience (MC2), Chalmers University of Technology, S-412 96 G\"oteborg, Sweden}
	
	\date{\today}
	
	\begin{abstract}
We investigate heat circulators where a phase coherent region is contacted by three leads that are either normal- or superconducting. A magnetic field, and potentially the superconducting phases, allow to control the preferential direction of the heat flow between the three-different temperature-biased contacts.  The main goal of this study is to analyze the requirements for heat circulation in non-ideal devices, in particular focusing on sample-to-sample variations. Quite generally, we find that the circulation performance of the devices is good as long as only a few transport channels are involved. We compare the performance of circulators with normalconducting contacts to those with superconducting contacts and find that the circulation coefficient are essentially unchanged. 
	\end{abstract}

	\maketitle

\section{Introduction}

With the miniaturization of electronic circuits and the possibility of fabricating devices operating at the nanoscale, control and management of heat flows~\cite{li2012} is becoming increasingly important. On the one hand, the performance of nanodevices can critically depend on ultracold temperatures~\cite{Giazotto2006Mar}. Therefore, the ability to control heat flows is for instance very useful when developing microrefrigerators, making it possible to operate on-chip cooling~\cite{Giazotto2006Mar,pekola2004}. On the other hand, from a different perspective, heat control in circuits is a fundamental requirement, when heat itself is used for operations~\cite{paolucci2018,guarcello2018}, as in the field of coherent caloritronics~\cite{m_perez2014,Fornieri2017Oct,Hwang2020-review} that has recently attracted a lot of attention. 

In this context, it becomes important to design and analyze devices which provide versatile and tunable control over heat flows. Many such devices have been investigated so far, both theoretically~\cite{terraneo2002,li2004,segal2005,eckmann2006,zeng2008,ojanen2009,ruokola2009,wu2009,wu2009b,kuo2010,ruokola2011,gunawardana2010,martinez2013,fornieri2014,landi2014,sanchez2015,jiang2015,joulain2016,vicioso2018,bours2019,goury2019,giazotto2020} and experimentally~\cite{Chang2006,Schmotz2011,Martinez2015,Scheibner_2008,Partanen2018,Senior2020}, including thermal transistors~\cite{li2006,giazotto2014,sanchez2017,yang2019}, valves~\cite{strambini2014,Ronzani2018}, interferometers~\cite{Giazotto2012Dec,martinez2013b} and a large variety of thermal rectifiers. Another key element for heat management, which has been less studied, is represented by circulators: these are multiterminal systems that are able to steer the heat conduction in a preferential direction (e.g., from a given terminal to the next one only in the anticlockwise direction). With this motivation, a three-terminal heat current circulator has recently been suggested~\cite{hwang18}. Such a device works similarly as charge current circulators~\cite{Viola2014May,Bosco2017Feb,mahoney2017,chapman2017}, which are crucial for electronics. Beyond that, three-terminal structures for coherent charge current control are also of interest for quantum networks~\cite{Wu2002Jan,Strambini2009May}.

\begin{figure}[b]
\includegraphics[width=0.45\textwidth]{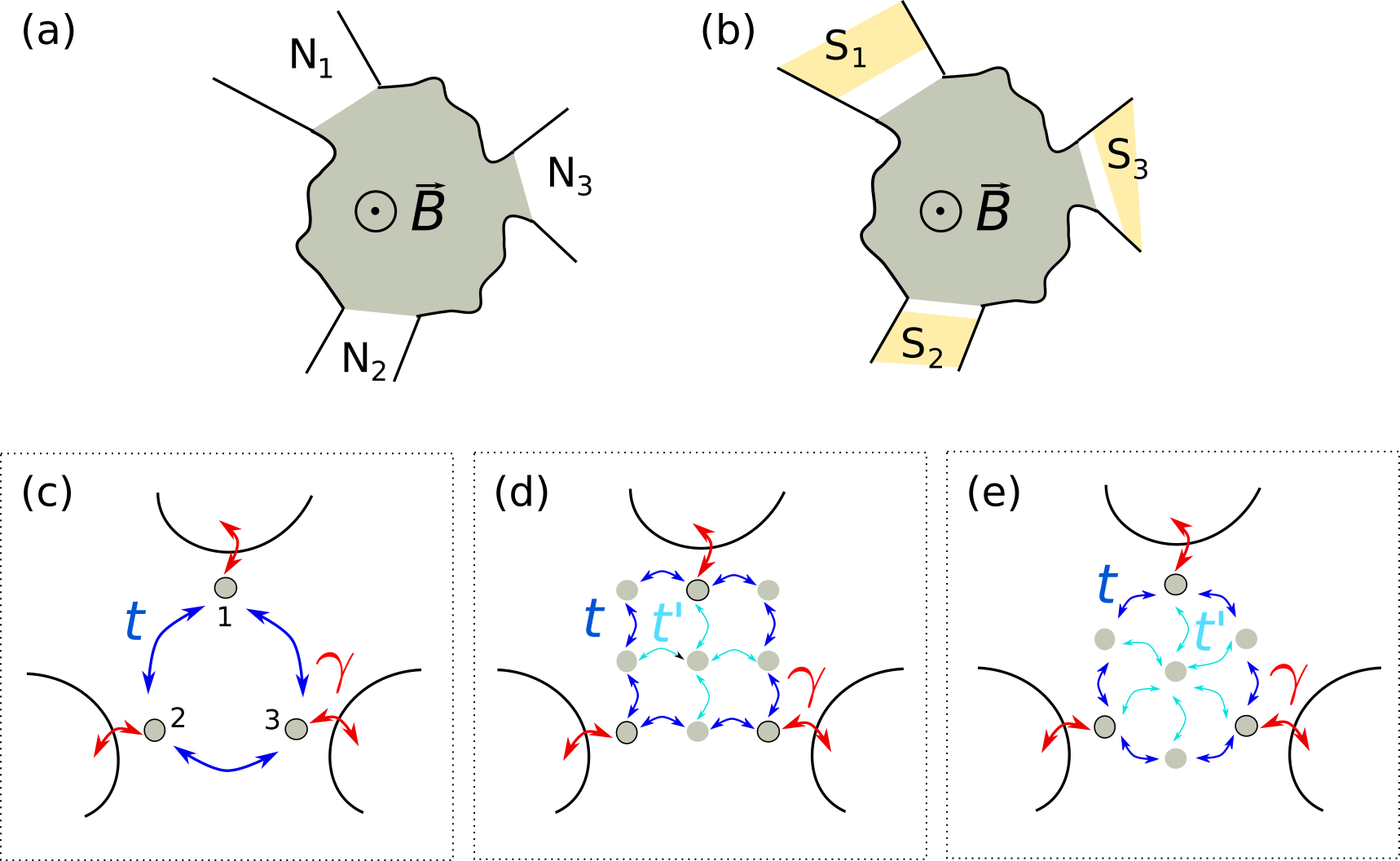}
\caption{Three-terminal conductors for heat circulation. (a) Generic setup consisting of a scattering region (gray) with three normal contacts denoted by $N_i$ and threaded by a magnetic field $B$. (b) The same setup with superconducting contacts. (c) Sketch of the simplest realization of the scattering region, made of three sites that are coupled to each other with hopping amplitude $t$ and to the contacts with $\gamma$. (d,e) Modifications of (c) allowing for return paths including different amounts of flux. Here, we have an additional hopping $t'$ to all sites not included in the external ``circle". These modified setups are addressed in Sec.~\ref{sec:nonideal}. \label{fig:setup}}
\end{figure}

The three-terminal device that has recently been suggested as a heat circulator consists of a ring-like structure with superconducting contacts, penetrated by a magnetic field, see Fig.~\ref{fig:setup}~(b). The superconducting system is of special interest for several reasons. It has been shown in recent years that heat currents carried by quasiparticles in superconducting devices can be coherently controlled via superconducting phase differences~\cite{Maki1965Dec,Zhao2003Aug,Giazotto2012Dec}. This has started the field of coherent caloritronics, reviewed in~\cite{Fornieri2017Oct}, where a circulator element can become part of the basic toolbox. From a fundamental perspective, superconducting phases give an additional control knob and heat can be circulated independently of charge currents. However, at the same time, heat control in general and heat circulation more specifically is equally of interest in normalconducting devices. Still, heat circulation in three-terminal normalconducting systems has, to our knowledge, not been considered so far. 

In this paper, we analyze three-terminal heat-current circulators from different perspectives. (1) We investigate the performance of heat circulation in a similar setup as in Ref.~\cite{hwang18} with normalconducting as well as with superconducting contacts, see Fig.~\ref{fig:setup}(a). This analysis actually shows that under certain conditions heat circulation can even be more effective in the absence of superconductivity. In contrast to the earlier proposal, where the gap was assumed to be constant, we fully include the self-consistent temperature-dependence of the superconducting gap. (2) The heat circulator of Ref.~\cite{hwang18} consists of a simplified setup with three central sites tunnel-coupled to each other and to the superconducting contacts,  penetrated by a homogeneous magnetic flux, see Fig.~\ref{fig:setup}(c).  We perform a detailed study on conductors deviating from the ideal ring-structure, see e.g. panel (d) and (e). A main consequence of this is that the enclosed flux varies with each trajectory that a particle can follow between the contacts. We show how this can lead to a deterioration and, in the most extreme cases, even to a full suppression of the circulation effect. We also discuss the conditions under which heat circulation is instead preserved. (3) We further investigate how realistic limitations can impact this ideal setting and analyze in detail the statistics of the circulating coefficient and its constituents. Previously, it has been shown how disorder can suppress phase-coherent control of heat flows~\cite{Virtanen2015Feb,Hajiloo2019Jun}. Here, we analyze the role of sample-to-sample variation of onsite energies and coupling constants in the three-sites structure and in the modified ones in Fig.~\ref{fig:setup}(d-e). In addition, we also consider as the central scattering region a chaotic cavity which we take as a model of an extended quantum system~\cite{alhassid2000}. Here, by using random matrix theory methods~\cite{beenakker-review}, we study the sample-to-sample variations of heat conductances and rectifications, in the same spirit of universal conductance fluctuations studies.

This analysis is important for experimental realizations of such circulators. However, most importantly, it gives detailed insights into the working principles of the circulator device. To improve the understanding of the device, we do not only study the circulation coefficient, but also analyze the heat conductances between terminals as well as rectification coefficients between pairs of terminals, separately. 

This paper is organized as follows. In Sec.\ \ref{sec:model} we lay out the general model and define the quantities we use to characterize a three-terminal device with either normal or superconducting contacts. The simplest realization of such a device [Fig.\ \ref{fig:setup}(c)] is investigated in Sec.\ \ref{sec:SvsN}, where normal and superconducting systems are compared. Then, more complicated setups, such as those in Fig.\ \ref{fig:setup}(d--e) and also a chaotic scattering region, are studied in Sec.\ \ref{sec:nonideal}. Finally, in Sec.\ \ref{sec:conclusions} we draw conclusions. Two Appendices are dedicated to technical and/or complementary details.

\section{Three-terminal conductors for heat circulation}\label{sec:model}

We consider a general three-terminal device, composed of three contacts, connected to a central conducting region (see Fig.~\ref{fig:setup}). Each terminal has a temperature $T_i$ $(i=1,2,3)$ and an electrochemical potential $\mu_i$. Our goal is to characterize the heat circulation properties of such a device, in the case where temperature biases (but no voltage ones) are present in the system. These transport properties depend on the details of the central region which is connected to the terminals and, in the framework of a scattering theory approach~\cite{Blanter2000,nazarov_book}, are determined by a scattering matrix. In contrast to Refs.~\cite{Spilla2014Apr,Spilla2015Jun}, we assume the central, ring-shaped device to be small with respect to the quasiparticle coherence length, guaranteeing coherent heat current control over the \textit{entire} central structure. As Figs.~\ref{fig:setup}(a-b) illustrate, we consider both the case where the terminals are normal metals and the one where they are superconductors. The main aspects distinguishing the two scenarios are that in the superconducting system heat and charge transport are very different and that the superconducting phases yield additional control parameters to tune the heat circulation. However, many of the properties we are interested in also occur in a normalconducting system and this is why we start with this case. In the remainder of this section we introduce the model describing the simplest realization of the scattering region [Fig.~\ref{fig:setup}(c)].

\subsection{Normalconducting heat circulator}\label{sec:N_model}

The most basic setup in which heat circulation is possible is a ring that can be modelled by a simple three-sites conductor~\cite{hwang18}. Any trajectory starting and ending at the same terminal encloses a magnetic flux $n\Phi$ which is an integer multiple ($n\in\mathbb{Z}$) of the magnetic flux $\Phi=\int_\text{ring}\!d^2r\, B$, given by the surface integral of the magnetic field $B$ (perpendicular to the plane) inside the ring.
For more complex structures, the enclosed flux depends on the details of the trajectories. This, including the extreme case, where the conductor is a chaotic cavity without any preferred trajectories, will be explored in Sec.~\ref{sec:nonideal}.

The setup represented in Fig.~\ref{fig:setup}(a), together with (c), is a normalconducting version of what has been investigated in Refs.~\cite{hwang18,meyer17}. It consists of three sites (labeled by $i=1,2,3$) with Hamiltonian $W$, that are connected by a hopping amplitude $\gamma$ to their respective lead. The Hamiltonian $W$ contains on-site energies $\epsilon_i$ on the diagonal and hopping amplitudes $\tilde{t}_{ij}$ (from $j$ to $i$) elsewhere. 
In the presence of an external magnetic field the hopping amplitudes have the form
$\tilde{t}_{ij}=t_{ij}e^{i\theta_{ij}}$, where $t_{ij}$ can be taken real and symmetric, while the phases $\theta_{ij}=-\frac{ie}{\hbar}\int_{j}^{i}\mathbf{A}\cdot d\boldsymbol{\ell}$ ($e>0$ is the elementary charge) encode the effect of the magnetic field. For any closed path, a particle thus gains a phase shift which is proportional to the flux enclosed by that path. In particular, one has the constraint $\theta_{13}+\theta_{32}+\theta_{21}=2\pi\Phi/\Phi_0\equiv\alpha$, with $\Phi_0=h/e$ the flux quantum.
Quite generally, for such a system the scattering matrix $\mathcal{S}(E)
$ at energy $E$ is obtained in terms of $W$ as (see Appendix \ref{app:S-derivation} for details)
\begin{equation}
    \mathcal{S}(E)=\left[\left(1-\frac{iE}{\Gamma}\right)\mathbbm{1}_3+\frac{i {W}}{\Gamma}\right]^{-1}\left[\left(1+\frac{iE}{\Gamma}\right)\mathbbm{1}_3-\frac{i{W}}{\Gamma}\right],
    \label{eq:S-normal}
\end{equation}
where $\Gamma=\pi\gamma^2\nu$ and $\nu$ is the density of states in the normal leads.
Considering the geometry in Fig.\ \ref{fig:setup}(c), we explicitly have
\begin{equation}
    W=\begin{pmatrix}
    \epsilon_1 & t_{12}e^{-i\alpha/3} & t_{13}e^{i\alpha/3}\\
    t_{12}e^{i\alpha/3} & \epsilon_2 & t_{23}e^{-i\alpha/3}\\
    t_{13}e^{-i\alpha/3} & t_{23}e^{i\alpha/3} & \epsilon_3
    \end{pmatrix}\,.
    \label{eq:W-3ring}
\end{equation}
Modifications of $W$, describing the setups in Fig.~\ref{fig:setup}(d-e), are introduced in section~\ref{sec:multi-flux}. Note that the scattering matrix obtained according to Eq.\ \eqref{eq:S-normal} is $2\pi$-periodic in the dimensionless  flux $\alpha$.

\subsection{Superconducting heat circulator}\label{sec:S_model}

In this subsection, we address the setup sketched in Fig.\ \ref{fig:setup}(b), where the three contacts are in a superconducting state. Each of them is therefore characterized by the (real-valued) magnitude of the gap $\Delta_i$ and a superconducting phase $\varphi_i$. In order to have a meaningful comparison with a normalconducting system introduced before, we self-consistently take into account the temperature dependence $\Delta_i(T_i)$ of the gap.
The scattering matrix $\mathcal{S}^{\text{sc}}$ of this three-terminal superconducting device can be expressed in terms of the scattering matrix $\mathcal{S}$ of the central region as~\cite{Beenakker1991Dec, nazarov_book}
\begin{equation}
    \mathcal{S}^\text{sc}=r+\tau'\mathcal{S}_0(\mathbbm{1}_6-r'\mathcal{S}_0)^{-1}\tau\,.
    \label{eq:S-sc}
\end{equation}
Here, the scattering at energy $E$ of electrons and holes at the central region is given by
\begin{equation}
    \mathcal{S}_0(E)=\begin{pmatrix}
    \mathcal{S}(E) & 0\\
    0 & {\mathcal{S}}^*(-E)\ .
    \end{pmatrix}
\end{equation}
The reflection (transmission) matrices $r,r'$ ($\tau,\tau'$) of the ideal normal-superconducting interfaces are given by~\cite{blonder82}
\begin{subequations}
\begin{align}
r_{eh}&=r_{he}=-\text{diag}\left(\left\{v_i u_i^{-1}\right\}\right)\,,\\
r'_{eh}&=\left(r'_{he}\right)^*=\text{diag}\left(\left\{e^{i\varphi_i}v_i u_i^{-1}\right\}\right)\,,\\
\tau_{ee}&=\left(\tau_{ee}'\right)^*={\tau}_{hh}'=\left(\tau_{hh}\right)^*\notag\\
&=\text{diag}\left(\left\{e^{i\varphi_i/2}u_i^{-1}\sqrt{u_i^2-v_i^2}\right\}\right)\,,
\end{align}
\end{subequations}
in the Andreev approximation. We define
\begin{equation}
    u_i^2=\frac{1}{2}\left(1+\frac{\sqrt{E^2-\Delta_i^2}}{E}\right)=1-v_i^2
\end{equation}
and $\text{diag}$ denotes a diagonal matrix.
With these ingredients, $\mathcal{S}^\text{sc}$ in Eq.~\eqref{eq:S-sc} fully describes the quasiparticle scattering processes at energy $E>\max_i(\Delta_i)$.

\section{Heat transport Observables}

Temperature differences between the three contacts lead to heat currents between them. Here, we are interested in controlling the magnitude of heat currents and the preferential direction in which they can circulate between the terminals.

\subsection{Heat current operator} 

The starting point is  the operator for the heat current into reservoir $i$ (see e.g. Ref.~\cite{moskalets-book})
\begin{align}\label{eq_operator}
    \hat{J}_i = &\frac1h\int_{E_\text{min}}^\infty dE'\int_{E_\text{min}}^\infty dE\sum_{n=e,h} \left(\frac{E+E'}{2}-\mu_i\right)\nonumber\\
    & \times\left[\hat{b}_{i,n}^\dagger(E)\hat{b}_{i,n}(E')-\hat{a}_{i,n}^\dagger(E)\hat{a}_{i,n}(E')\right]
\end{align}
and its expectation value $\langle \hat{J}_i\rangle =J_i$. Here, we measure all energies with respect to the electrochemical potentials $\mu_i\equiv0$ which we assume to be equal for all $i$. The annihilation operators $\hat{a}_{i,n}$ for incoming fluxes  are connected to operators $\hat{b}_{i,n}$ for outgoing ones by the (elastic) scattering matrices given in the previous section. The subscript $n$ indicates electron- and hole-like contributions to transport. To keep the notation simple, we write Eq.~(\ref{eq_operator}) for the single-channel case; for multi-channel cases, such as treated in Sec.~\ref{sec_multi}, the sum over $n$ needs to be extended to count channels in the contacts too. For a normalconducting system, the integral starts at $E_\text{min}=0$, while the lower bound is $E_\text{min}=\Delta$ in the superconducting case. Here, $\Delta=\Delta(T_i)=\text{max}_i(\Delta_i(T_i))$ is the magnitude of the largest gap $\Delta_i$ in the problem at given temperatures $T_i$.

\subsection{Heat conductance} 

To characterize the heat transport of the devices shown in Fig.~\ref{fig:setup}, we consider the linear-response heat conductances in the presence of small temperature differences. A standard linear-response calculation for $J_i=\sum_j\kappa_{ij}|_{T_{i,j}\equiv T}\delta T_j$ yields the heat conductance matrix with elements $\kappa_{ij}\equiv\kappa_{ij}|_{T_{i,j}\equiv T}$. For the normalconducting case, one finds, taking together electron- and hole contributions,
\begin{equation}
\kappa_{ij}=-\frac{1}{hT}\int_{-\infty}^{\infty} dE\,E^2\,f'(E)\left(\delta_{ij}-\left|{\mathcal{S}_{ij}}\right|^2\right)\ .
\label{eq:kappa-normal}
\end{equation}
Here, $f(E)=[1+\exp(E/k_\text{B}T)]^{-1}$ is the Fermi function at temperature $T$.
For the system with superconducting contacts, we  obtain instead
\begin{equation}
    \kappa_{ij}^{\text{sc}}=-\frac{1}{hT}\int_{\Delta}^{\infty} \!\!dE\,E^2 f'(E) \left[2\delta_{ij}-\text{Tr}\left({\mathcal{S}_{ij}^\text{sc}}^\dagger\mathcal{S}^\text{sc}_{ij}\right)\right]\,.
\end{equation}
The trace appearing in this case is to be taken over the electron-hole degrees of freedom.

\subsection{Heat rectification and circulation}

It has been shown in Ref.~\cite{hwang18} that it is possible to achieve a situation where the heat preferentially flows in a given direction by tuning the magnetic flux and/or imposing a superconducting phase bias among the terminals. 
In linear response, this requires that the heat conductances present a rectification effect with $\kappa_{ij}\neq\kappa_{ji}$ for $i\neq j$. To quantify this effect, we introduce the rectification coefficient
\begin{align}
    \label{eq:rect}
    \mathcal{R}_{ij}=\frac{\kappa_{ij}-\kappa_{ji}}{\kappa_{ij}+\kappa_{ji}}\ .
\end{align}
The heat circulation is quantified via the coefficient
\begin{equation}\label{eq:C}
    \mathcal{C}=\frac{\kappa_{13}\kappa_{32}\kappa_{21}-\kappa_{12}\kappa_{23}\kappa_{31}}{\kappa_{13}\kappa_{32}\kappa_{21}+\kappa_{12}\kappa_{23}\kappa_{31}}\,.
\end{equation}
It takes the value $\mathcal{C}=+1$ for a perfect counterclockwise circulation and $\mathcal{C}=-1$ for a perfect clockwise one, which arise for the limiting cases, where either $\kappa_{ij}=0$ or $\kappa_{ji}=0$. In the superconducting system, the definitions for $\mathcal{R}^\text{sc}$ and $\mathcal{C}^\text{sc}$ are identical to the previous ones, simply replacing each $\kappa_{ij}$ with $\kappa^\text{sc}_{ij}$.

It is instructive to understand how the circulation and rectification coefficients $\mathcal{C}$ and $\mathcal{R}$ are related to each other. In the fully symmetric case, where all anti-clockwise heat conductances are equal to $\kappa_{\overline{\text{cw}}}$ and all the clockwise ones are equal to $\kappa_{\text{cw}}$, the rectification and circulation are given by
\begin{align}
    \mathcal{R}=\frac{
                        \kappa_{\overline{\text{cw}}}-\kappa_\text{cw}
                     }{
                        \kappa_{\overline{\text{cw}}}+\kappa_\text{cw}
                      } \ \ \text{and} \ \ 
    \mathcal{C}=\frac{
                        \kappa^3_{\overline{\text{cw}}}-\kappa_\text{cw}^3
                      }{
                        \kappa^3_{\overline{\text{cw}}}+\kappa_\text{cw}^3
                      } \ .
\end{align}
The relation between the two coefficients is more direct in the case of weak rectification, $\kappa_{ij}-\kappa_{ji}\ll\kappa_{ij}+\kappa_{ji}$, where we find
\begin{align}
    \mathcal{C}&\approx\mathcal{R}_{12}+\mathcal{R}_{23}+\mathcal{R}_{31}. 
    \end{align}

\section{Linear-response heat circulation in an ideal three-sites setup}
\label{sec:SvsN}

In this section, we show the results of our analysis for the ideal three-sites setup of Fig.~\ref{fig:setup}(c), starting with the normalconducting system and then comparing its performance with the superconducting heat circulator.

\subsection{Normalconducting contacts}\label{sec:linear_normal}

\begin{figure}[t]
\includegraphics[width=\columnwidth]{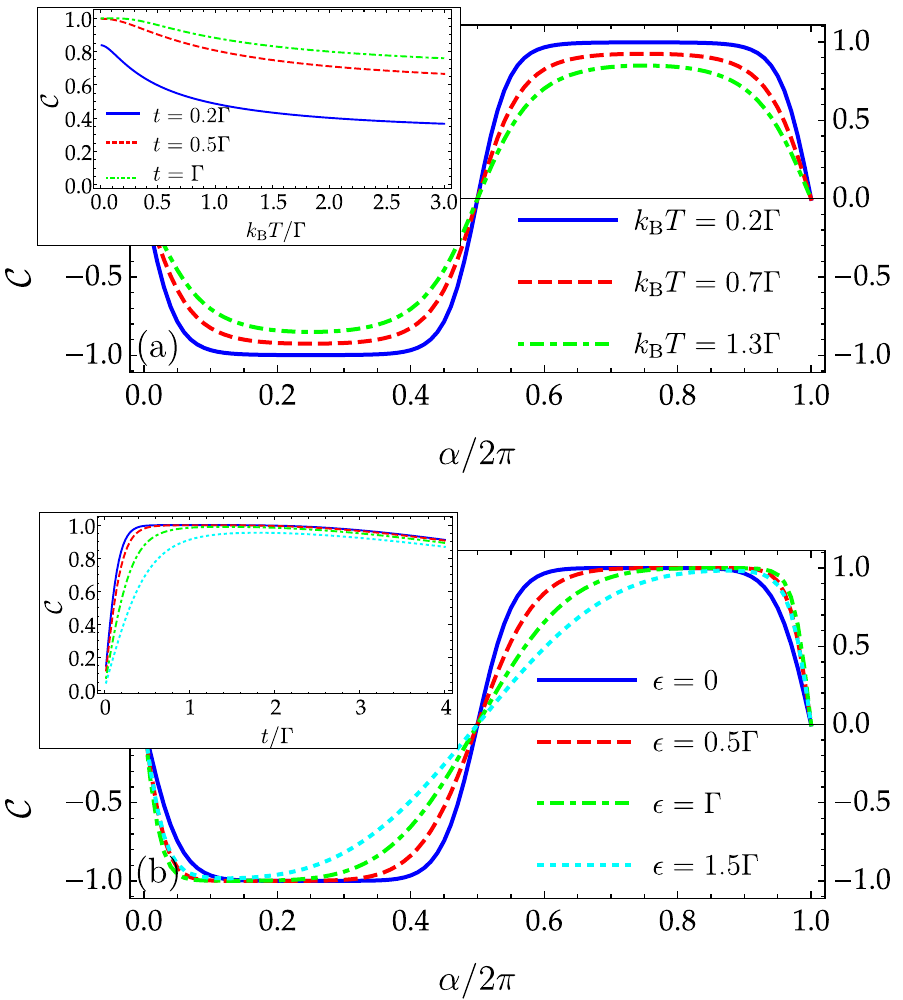}
\caption{Circulation coefficient $\mathcal{C}$ for the basic three-sites ring. In panel (a) as a function of the dimensionless magnetic flux $\alpha$ for a hopping amplitude $t=\Gamma$ and different temperatures (inset: as a function of temperature for different hopping amplitudes and $\alpha=3\pi/2$). (b) Low-temperature limit, where $\kappa_{ij}$ can be evaluated from Eq.~\eqref{eq:kappa-normal-simplified}. Here, we plot again $\mathcal{C}$ vs $\alpha$ for different on-site energies (inset: $\mathcal{C}$ vs $t/\Gamma$ for $\alpha=3\pi/2$ and the same values of $\epsilon$ as in the main plot).
\label{fig:C-alpha-3ring}}
\end{figure}

We consider the specific setup combining Fig.~\ref{fig:setup}(a) with (c), for which the model has been introduced in Sec.~\ref{sec:N_model}. Its properties are particularly simple in the symmetric situation with equal on-site energies $\epsilon_i=\epsilon$ and hopping amplitudes $t_{ij}=t$; we will address this simple case in the present section. Then, all heat conductances in the same direction (clockwise/anticlockwise) are equal to each other. The circulation coefficient, obtained from Eqs.\ \eqref{eq:kappa-normal} and \eqref{eq:C}, is shown as a function of the magnetic flux in Fig.\ \ref{fig:C-alpha-3ring}(a) and as a function of temperature (inset of (a)). Concerning the dependence on the magnetic flux, the ideal ring shape of the system constrains $\mathcal{C}$ to vanish at $\alpha=0,\,\pi,\,2\pi$, meaning that for these fluxes it is equally probable for the heat current to circulate in both directions. Moreover, the circulation coefficient is antisymmetric about the point $\alpha=\pi$ and is maximal at $\alpha=3\pi/2$ for a large range of parameters. As far as temperature effects are concerned, Fig.\ \ref{fig:C-alpha-3ring}(a) shows that by increasing $T$, the circulation coefficient is typically reduced. The typical scale on which the effect of a finite temperature becomes important is $k_\text{B}T\sim\Gamma$.

This means in particular, that in the regime $k_\text{B}T\ll\Gamma$ the temperature dependence of the circulation coefficient (and of the underlying heat conductances) is negligible. The reason for this is that the scattering matrix $\mathcal{S}$ can in this limit be taken as energy-independent, setting $E=0$ in Eq.\ \eqref{eq:S-normal}. Then, Eq.\ \eqref{eq:kappa-normal} reduces to
\begin{equation}
    \kappa_{ij}=\kappa_0\left(\left|{\mathcal{S}_{ij}}\right|^2-\delta_{ij}\right)\,,
    \label{eq:kappa-normal-simplified}
\end{equation}
where $\kappa_0=\pi^2k_\text{B}^2T/(3h)$ is the thermal conductance quantum.\footnote{Notice that this approximation was indeed used in Ref.~\cite{hwang18} where the superconducting circulator was proposed. As we discuss in the following, this is a point to be careful about, as the presence of the superconductor in general further enhances the energy-dependent features of the scattering matrix.}
The scattering matrix elements at $E=0$, obtained from \eqref{eq:S-normal}, have the simple form
\begin{equation}
\begin{split}
    \mathcal{S}_{31}&=\mathcal{S}_{12}=\mathcal{S}_{23}=\frac{2e^{2i\alpha/3}\bar{t}(\bar{t}+iye^{-i\alpha})}{2i{\bar{t}}^3\cos\alpha-y(3{\bar{t}}^2+y^2)}\,,\\
    \mathcal{S}_{13}&=\mathcal{S}_{21}=\mathcal{S}_{32}=\frac{2e^{-2i\alpha/3}\bar{t}(\bar{t}+iye^{i\alpha})}{2i{\bar{t}}^3\cos\alpha-y(3{\bar{t}}^2+y^2)}\,,\\
    \mathcal{S}_{jj}&=\frac{{\bar{t}}^2(1+3i\bar{\epsilon})-y^2y^*-2i{\bar{t}}^3\cos\alpha}{2i{\bar{t}}^3\cos\alpha-y(3{\bar{t}}^2+y^2)}\,,
\end{split}
\end{equation}
where $\bar{\epsilon}=\epsilon/\Gamma$, $\bar{t}=t/\Gamma$ and $y=1+i\bar\epsilon$. Note, that in this special case,  heat and charge circulation are the same, as the thermal and electrical conductances are related by the Wiedemann-Franz law.

We now show how $\mathcal{C}$ depends on $\epsilon$ and $t$, in this low-temperature regime, $k_\text{B}T\ll\Gamma$.
In Fig.\ \ref{fig:C-alpha-3ring}(b),  $\mathcal{C}$ is shown as a function of $\alpha$ for different on-site energies $\epsilon$ and the inset shows the dependence on $t$. We see that the optimal hopping amplitude is $t=\Gamma$ but a close-to-maximal circulation coefficient is also found for a quite large range of values around $t=\Gamma$. In addition, Fig.\ \ref{fig:C-alpha-3ring}(b) shows that the circulation coefficient is not very sensitive to $\epsilon$: a deviation of the on-site energy from $\epsilon=0$ makes the curves more asymmetric, slightly shifting the value of $\alpha$ at which the maximum circulation coefficient is reached. Overall, as also shown in the inset, the effect of increasing $|\epsilon|$ is to reduce the circulation coefficient.

\subsection{Superconducting contacts}\label{sec:linear_super}

We now compare the results for the normalconducting device, analyzed above, with those of a superconducting circulator, introduced in Sec.\ \ref{sec:S_model}. Before starting the discussion, it is worth noting that, in the absence of phase biases, also the superconducting device is completely symmetric and therefore all clockwise heat  conductances are equal to each other (idem for the anticlockwise heat conductances). We will first consider precisely this case and show the effect of phase biases later on. Moreover, in order to keep the discussion as simple as possible, we assume from now on that the three superconductors have the same gap amplitude $\Delta=\Delta_1=\Delta_2=\Delta_3$, the temperature dependence of which is calculated self-consistently.

\begin{figure}[b]
\includegraphics[width=\columnwidth]{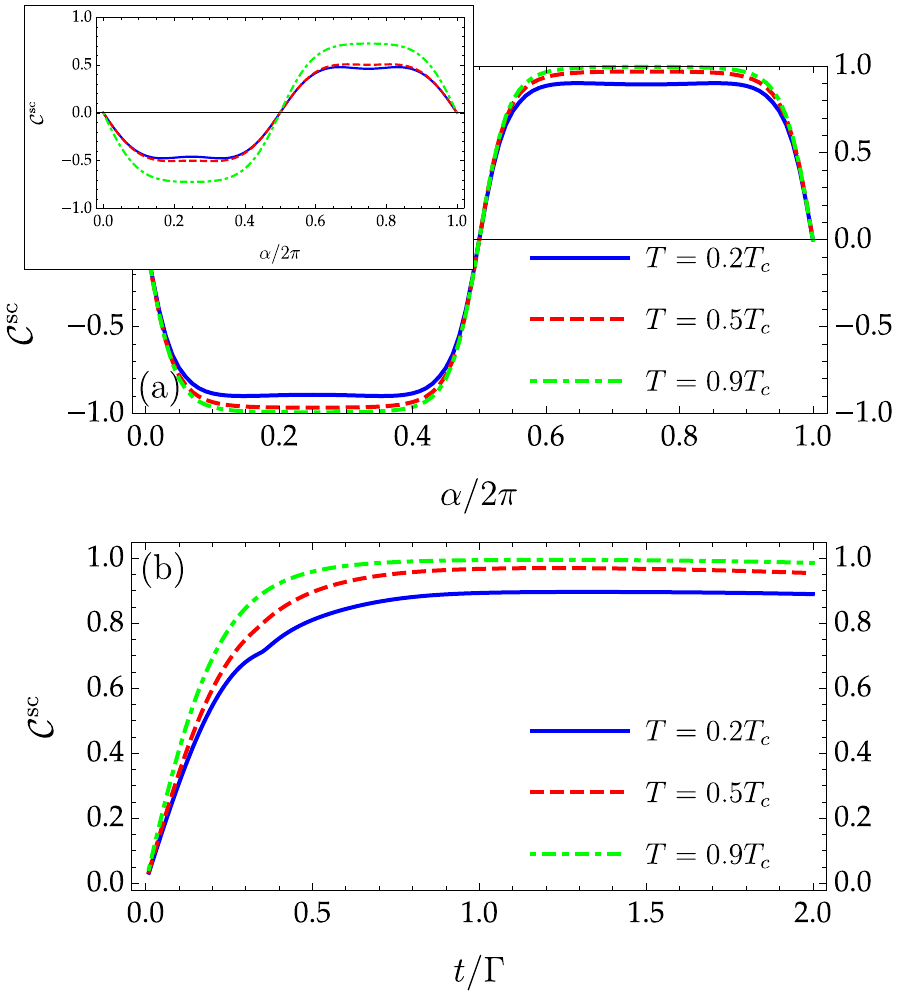}
\caption{(a) Circulation coefficient $\mathcal{C}^\text{sc}$ for the superconducting system with critical temperature $T_\text{c}$ as a function of the magnetic flux $\alpha$ for $\epsilon=0$ and for a hopping amplitude $t=\Gamma=5k_\text{B}T_c$. Inset: the same, but with $\Gamma=k_\text{B}T_c$. (b) $\mathcal{C}^\text{sc}$ as a function of $t$, for $\alpha=3\pi/2$, $\epsilon=0$ and $\Gamma=5k_\text{B}T_c$. In both panels, no phase bias is applied between the superconducting contacts.
\label{fig:Csc-vs-alpha}}
\end{figure}

We show in Fig.\ \ref{fig:Csc-vs-alpha}(a) $\mathcal{C}^\text{sc}$ as a function of $\alpha$ for $\epsilon=0$, different temperatures and $t=\Gamma=5k_\text{B}T_c$, where $T_c$ is the critical temperature of the superconductor. Similarly to the normal system, in this low-temperature regime, it is possible to reach a close-to-maximal clockwise $(\mathcal{C}^\text{sc}=-1)$ or counterclockwise $(\mathcal{C}^\text{sc}=+1)$ circulation. 
Hence, as for the normal circulator, the most favorable regime for a good circulation coefficient is the low-temperature regime. Again this regime is fixed by $k_\text{B}T\ll\Gamma$, where now $T$ should at the same time not exceed the critical temperature $T_\text{c}$ for the system to be in the superconducting state. Qualitatively, in this regime we find the same result as in Ref.\ \cite{hwang18}, however at modified parameters and with small changes due to the self-consistent evaluation of the temperature-dependence of the gap amplitude $\Delta$.

When reducing $\Gamma$ with respect to $T$ and $T_c$, $\mathcal{C}^{\text{sc}}$ is typically reduced: an example of such a trend is shown in the inset, where $\Gamma=k_\text{B}T_c$. This considerable temperature-dependence is due to the fact that the energy dependence of the scattering matrix is enhanced compared to the one of the normalconducting system by the presence of the superconducting gap.

As further features, we observe that the value of $\alpha$ at which the maximal circulation coefficient $|\mathcal{C}^\text{sc}|$ is reached in general depends on the interplay between the parameters $t$, $T$ and $\Gamma$. In many cases, though, the maximum is found at $\alpha=3\pi/2$, as in Fig.\ \ref{fig:Csc-vs-alpha}(a). In addition, Fig.~\ref{fig:Csc-vs-alpha}(b) shows that the maximal circulation coefficient at $\alpha=3\pi/2$ is achieved for a hopping amplitude close to $t=\Gamma$ (here, again, the exact value of the maximum as a function of $t$ slightly depends on temperature). This plot is specific to the choice $\Gamma/k_\text{B}T_c\gg1$: at lower ratios, the behavior is quite different (not shown).
\begin{figure}[t]
\includegraphics[width=0.9\columnwidth]{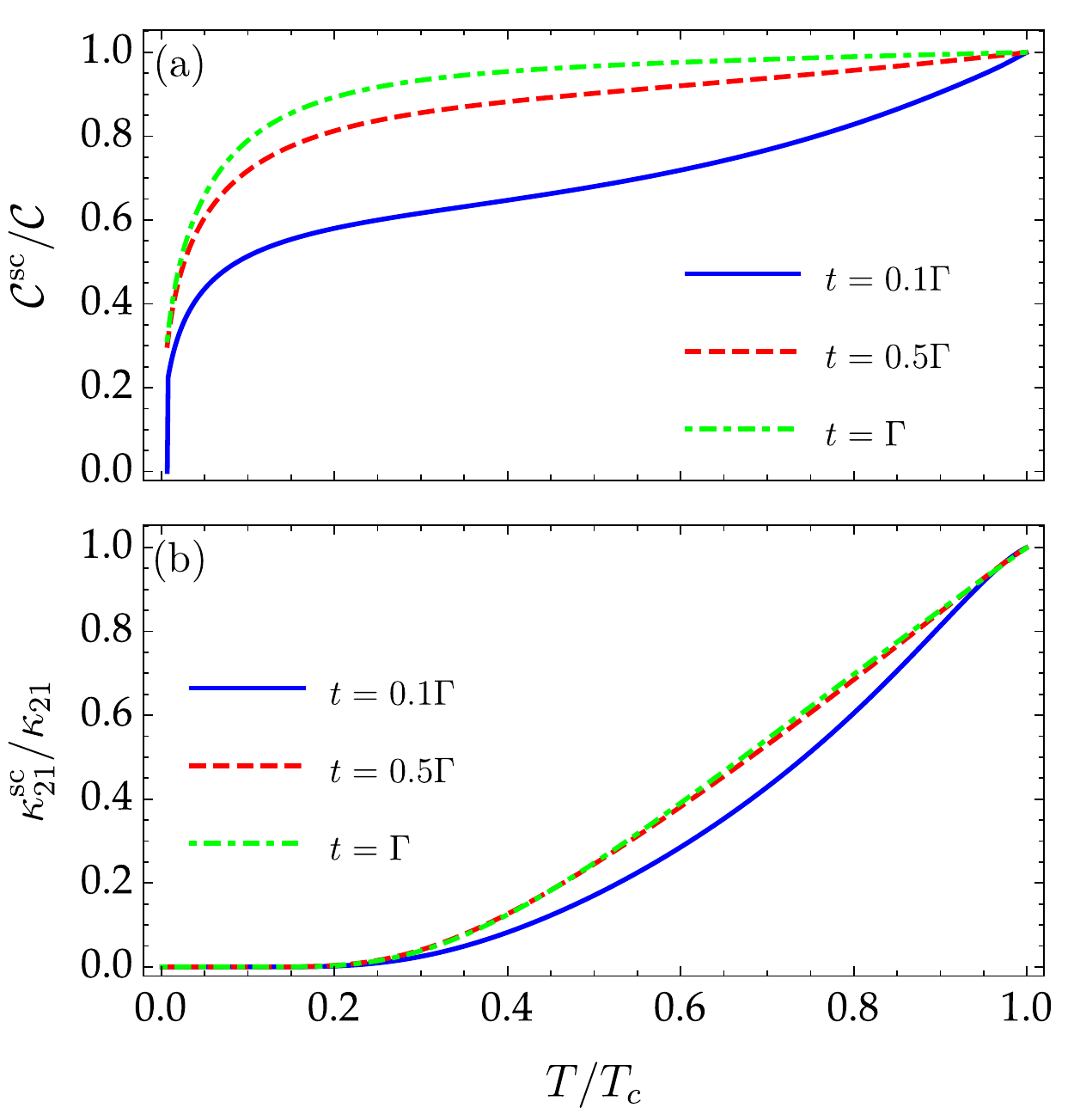}
\caption{Ratio of the circulation coefficients (a) and the heat conductances (b) between the superconducting and normal device as a function of temperature. We have set the phases of the superconductors to be equal and chosen $\alpha=3\pi/2$, $\epsilon=0$ and $\Gamma=5k_\text{B}T_c$.
\label{fig:Cs-over-Cn}}
\end{figure}

Comparing  Fig.\ \ref{fig:C-alpha-3ring} and \ref{fig:Csc-vs-alpha} only reveals small differences between the normal- and superconducting systems. It is natural to ask whether the superconducting device offers some advantages with respect to the normal one. Let us start with the case where no phase biases are imposed among the superconducting terminals and compare the performances of the normal- and superconducting devices, which we present in Fig.\ \ref{fig:Cs-over-Cn}(a). Here, we show the ratio between the circulation coefficients $\mathcal{C}^\text{sc}$ and $\mathcal{C}$ at $\alpha=3\pi/2$, where we have the maximal counterclockwise heat circulation for most values of $t$. We see that the normal system usually performs better than the superconducting one, at any temperature.\footnote{When $T=T_\text{c}$, the two circulation coefficients are equal, since superconductivity is suppressed.} In Fig.\ \ref{fig:Cs-over-Cn} we chose $\Gamma/k_\text{B}T_c=5$. We have verified that only for much larger ratios (over 30), it is possible that $\mathcal{C}^\text{sc}>\mathcal{C}$ for hoppings $t\sim 0.1\Gamma$. 
In addition, one should be aware that the heat conductance in the superconducting device is typically smaller than that of the normal one, as shown in Fig.\ \ref{fig:Cs-over-Cn}(b). This means that the \textit{amount} of circulated heat is typically larger in the normalconducting compared to the superconducting setup. We can therefore conclude that, in the absence of phase biases, there is no specific advantage of the superconducting  with respect to the normalconducting device.

\begin{figure}[t]
\includegraphics[width=0.9\columnwidth]{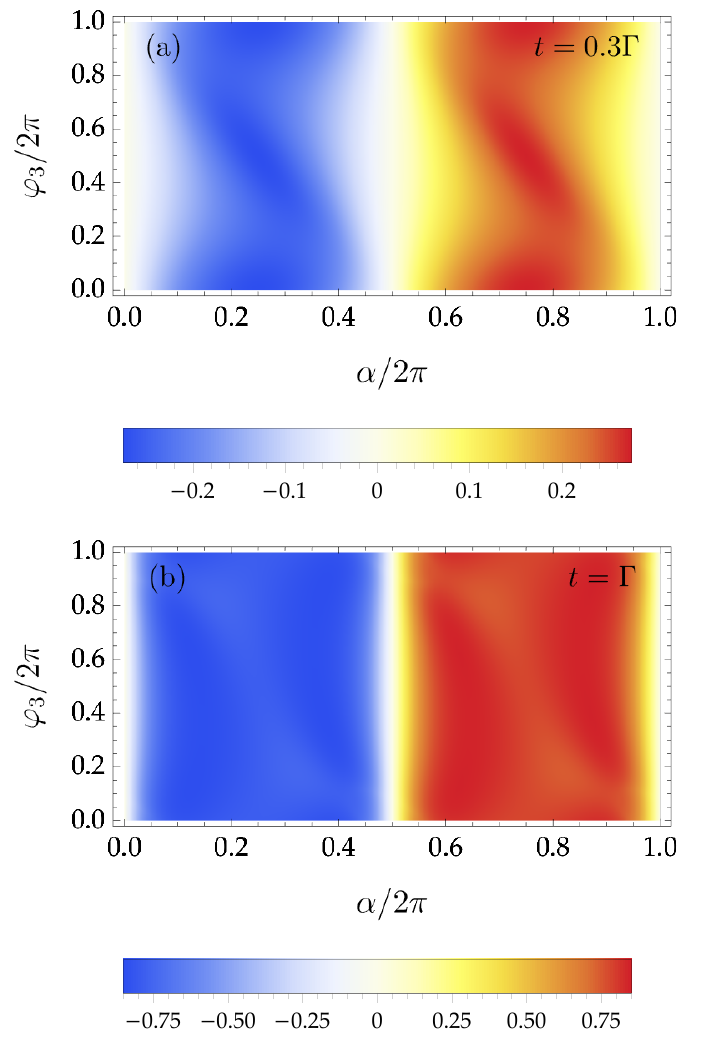}
\caption{Circulation coefficient $\mathcal{C}^\text{sc}$ as a function of $\alpha$ and the superconducting phase $\varphi_3$ in terminal 3 (with respect to $\varphi_1=\varphi_2=0$) for two values of the hopping amplitude $t$. We have set $\epsilon=0$, $T=0.1T_\text{c}$ and $\Gamma=5k_\text{B}T_c$.
\label{fig:C-vs-flux-phase}}
\end{figure}

Nevertheless, the latter is more versatile as it offers an additional control parameter to tune the circulation, namely the possibility of imposing phase biases between different terminals. In Ref.~\cite{hwang18} it was shown that when the heat circulation is controlled by just imposing phase biases (and no magnetic flux), an opposite behavior compared to Fig.\ \ref{fig:Csc-vs-alpha}(b) emerges. That is, a higher circulation coefficient is reached at lower hopping amplitudes $t$. Here, we show the combined effect of phase bias and magnetic field: in Fig.\ \ref{fig:C-vs-flux-phase} we plot $\mathcal{C}^\text{sc}$ as a function of $\alpha$ and $\varphi_3$, fixing the other superconducting phases to $\varphi_1=\varphi_2=0$. As we can see, the highest circulation coefficient is reached when $t$ is close to  $\Gamma$, namely when the dependence of $\mathcal{C}^\text{sc}$ on the phase $\varphi_3$ is quite weak [Fig.\ \ref{fig:C-vs-flux-phase}(b)]. In contrast, at low hoppings the circulation is more sensitive to variations of the superconducting phase, albeit the overall circulation coefficient is smaller [Fig.\ \ref{fig:C-vs-flux-phase}(a)].

\section{Sample-to-sample variations}
\label{sec:nonideal}

Having described the behavior of the simplest possible setup, in this section, we show how non-ideal operational conditions affect the performance of the heat circulator. Focusing on the low-temperature regime, where the energy dependence of the scattering matrix in $\kappa$ can be neglected, see Eqs.~(\ref{eq:kappa-normal}) and (\ref{eq:kappa-normal-simplified}), we consider two other mechanisms leading to deviations from the ideal condition. A first ingredient is represented by random variations of the parameters of the model (hopping amplitudes and on-site energies), in order to see whether sample-to-sample variations limit the usefulness of the device. Secondly, we investigate what happens when the device does not have a ring-like structure, as shown in the modified setups in Fig.\ \ref{fig:setup}(d-e). Here, the main difference with the basic model in Fig.\ \ref{fig:setup}(c) is that a path starting and ending at the same terminal can include different fractions of the total magnetic flux penetrating the structure. To avoid unnecessary complications,  we restrict the analysis to the normalconducting case. This choice is motivated by the fact that superconducting terminals exhibit qualitatively the same behavior, as shown in the previous section.

\subsection{Variations of the hopping and onsite energies}

\begin{figure}[t]
\includegraphics[width=\columnwidth]{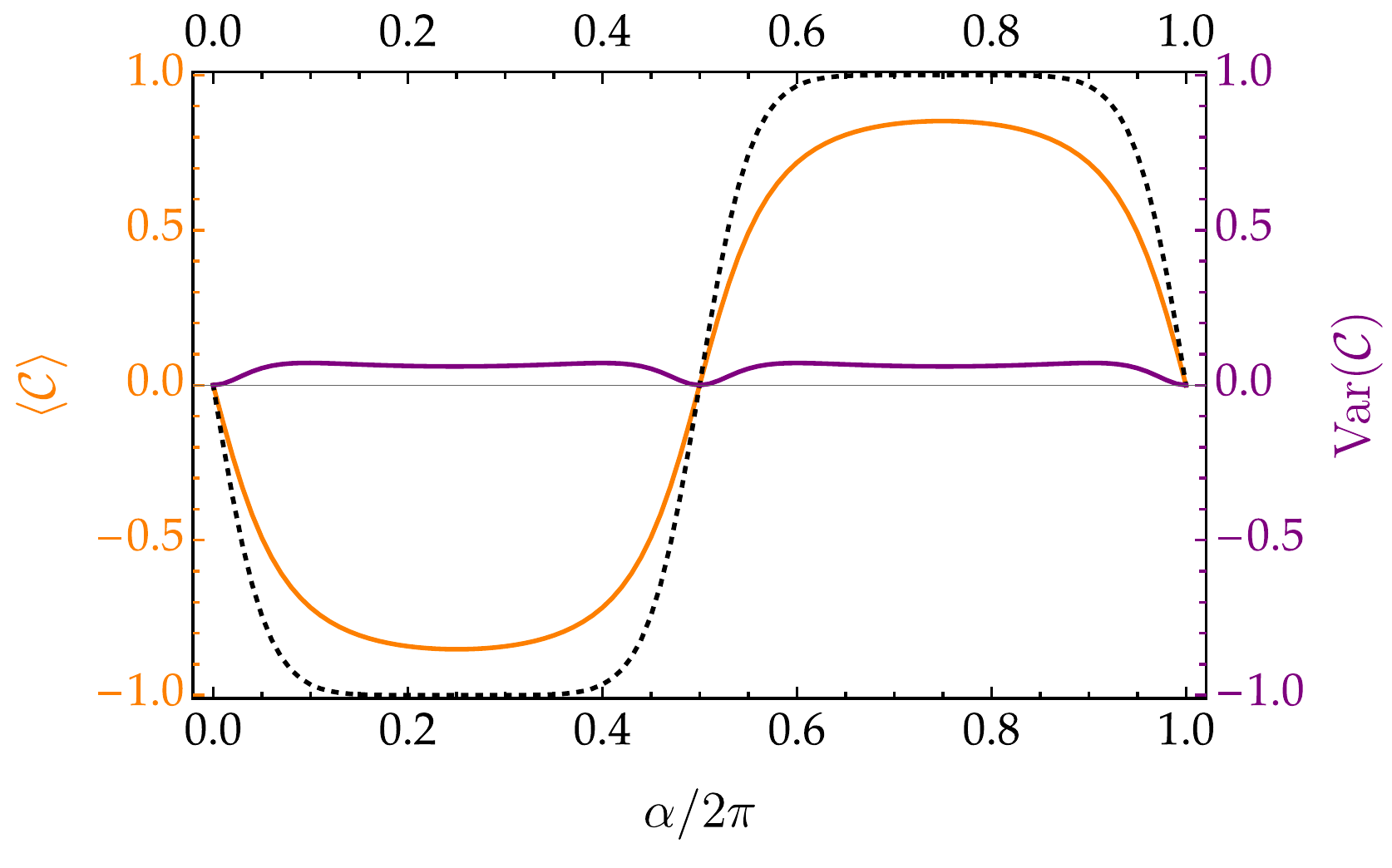}
\caption{Comparison between the ideal circulation coefficient for $t_{ij}=t=\Gamma$ and $\epsilon_i=\epsilon=0$ (dashed black line) and the ensemble-averaged circulation coefficient $\braket{\mathcal{C}}$ over $2000$ samples obtained with a random variation of the parameters in the coupling matrix  Eq.~\eqref{eq:W-3ring} (solid orange line), see main text. The variance of $\mathcal{C}$ is also shown (solid purple line).
\label{fig:C-3ring-random}}
\end{figure}

As previously shown, the circulation coefficient of the ideal three-sites system is not particularly sensitive to the variation of the hopping amplitude and on-site energy (see Fig.\ \ref{fig:C-alpha-3ring}). Therefore, $\mathcal{C}$ is also expected to be quite robust to \textit{sample-to-sample variations} of the model parameters. In Fig.~\ref{fig:C-3ring-random}, we show that this is indeed the case, a result that was already anticipated in Ref.~\cite{hwang18}. We do not restrict ourselves to the symmetric case $\epsilon_i=\epsilon$ and $t_{ij}=t$, but we consider the general coupling matrix in Eq.~\eqref{eq:W-3ring}.
In particular, we have taken $t_{ij}, \epsilon_i$ to be uniformly distributed, i.e.~$|t_{ij}-\braket{t_{ij}}|<\delta(t_{ij})/2$, with average $\braket{t_{ij}}=\Gamma$ and full width $\delta(t_{ij})=2\Gamma$ and $|\epsilon_i-\braket{\epsilon_i}|<\delta(\epsilon_i)/2$, with $\braket{\epsilon_i}=0$ and $\delta(\epsilon_i)=2\Gamma$. We find that  the reduction of the circulation coefficient is roughly only about $15\%$ (clearly by reducing the range of variations of the parameters, the performance of the device is even less affected). Furthermore, Fig.\ \ref{fig:C-3ring-random} shows that $\mathcal{C}$ varies only a little around its ensemble-average $\braket{\mathcal{C}}$. All these features are peculiar of the simple model of Fig.~\ref{fig:setup}(c); in the following, we show that with increasing complexity of the device also the impact of the sample-to-sample variations grows.

\subsection{Trajectory-dependent enclosed flux}
\label{sec:multi-flux}

As a next step, we investigate the impact of a modified structure of the central scattering region on the behavior of the heat circulator device. In particular, we consider the possibility that different paths starting and ending at a given terminal enclose a different magnetic flux. First, we consider a modification of the simple toy model that we obtain by adding extra lattice sites to the minimal model, as shown in Figs.~\ref{fig:setup}(d) and (e). Secondly, we consider a more realistic model of an extended central region that we describe by a chaotic cavity modelled by random scattering matrices. As we will see, while single realizations can still be tuned to act as heat circulators, these increasingly complicated systems have a considerable impact on the device performance.

\subsubsection{Generic central structure with multiple lattice sites}\label{sec:Multiring}
We consider a modified setup, as represented in Figs.\ \ref{fig:setup}(d) and (e): here, $\mathcal{M}$ additional sites (labeled by $\mu=a_1,\dots,a_\mathcal{M}$) are present in the scattering region, but they are not connected to any terminal. Thus, in Fig.\ \ref{fig:setup}(d) we have $\mathcal{M}=6$ and in Fig.\ \ref{fig:setup}(e) $\mathcal{M}=4$. In both cases, the extra central site is labelled with $\mu=a_1$. In this modified setup, the coupling matrix $\mathcal{W}$ describing the central region can be split into four blocks as
\begin{equation}\label{eq_schur}
    \mathcal{W}=
    \begin{pmatrix}
    \mathcal{W}^{11} & \mathcal{W}^{12}\\
    \mathcal{W}^{21} & \mathcal{W}^{22}
    \end{pmatrix}\,,
\end{equation}
where $\mathcal{W}^{11}$ characterizes the subspace of the three sites connected to the leads, $\mathcal{W}^{22}$ characterizes the subspace of the additional $\mathcal{M}$ sites, and cross couplings between sites $i=1,2,3$ and $\mu=a_1,\dots,a_\mathcal{M}$ are taken into account by the off-diagonal blocks $\mathcal{W}^{12}$ and $\mathcal{W}^{21}$. For this more general system, a formally identical result as in Eq.\ \eqref{eq:S-normal} is obtained, with a modified matrix $W$ given by the Schur complement $W=\mathcal{W}^{11}-\mathcal{W}^{12}{(\mathcal{W}^{22})}^{-1}\mathcal{W}^{21}$ (see App.\ \ref{app:S-derivation} for details).

Although the result is valid for any $\mathcal{W}$, we restrict the discussion to nearest neighbor coupling only, as shown in the sketch of Fig.\ \ref{fig:setup}(d) and (e). The phases associated with the transition from any site $\zeta$ to any of its neighbors $\xi$ are calculated according to $\theta_{\xi\zeta}=-\frac{ie}{\hbar}\int_{\zeta}^{\xi}\mathbf{A}\cdot d\boldsymbol{\ell}$, using the symmetric gauge $\mathbf{A}=B(-y,x,0)/2$. As for the basic three-sites model, the resulting scattering matrix is periodic in the normalized flux $\alpha$, although the periodicity is no longer $2\pi$, due to the more complex possible paths that can be followed between any two terminals. It depends on the detailed geometry of the system, as we discuss in the following.

\subsubsection{Hexagonal central structure}

\begin{figure}[t]
\centering
\includegraphics[width=\columnwidth]{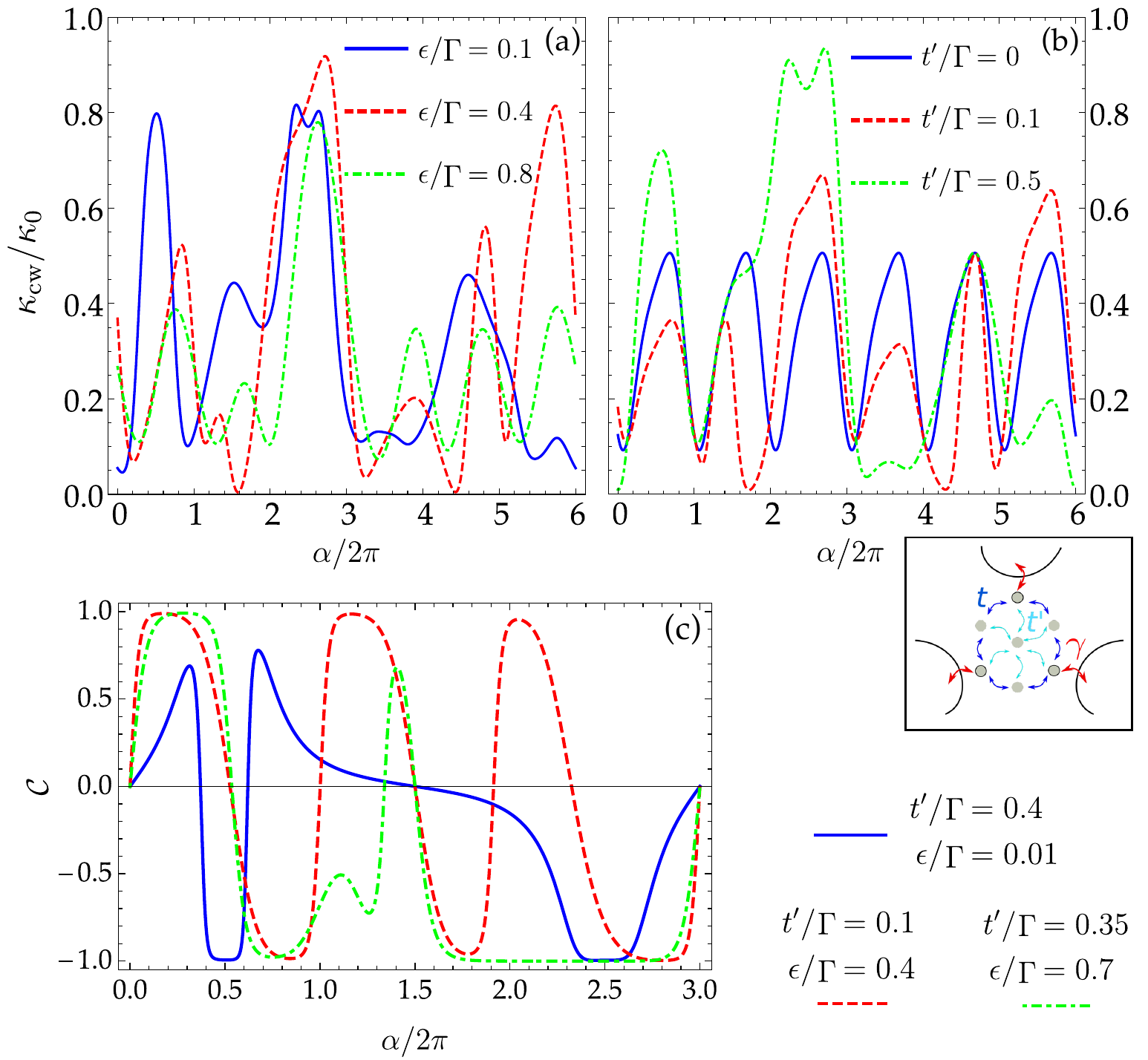}
\caption{(a--b) Heat conductance $\kappa_{\text{cw}}$ as a function of the magnetic flux for the setup depicted in Fig.~\ref{fig:setup}(e). In both panels the hopping to the leads is $t=0.5\Gamma$. We have set $\epsilon=0.2\Gamma$ in (a) and $t'=0.2\Gamma$ in (b). (c) Circulation coefficient for $t=0.5\Gamma$ and different values of $\epsilon$ and $t'$. Note that for all choices of parameters, the circulation $|\mathcal{C}|$ can reach almost unity by fine-tuning the magnetic flux $\alpha$.
\label{fig:kappa-7ring}}
\end{figure}

Let us start with the analysis of the ``hexagonal'' model in Fig.~\ref{fig:setup}(e). We initially consider equal on-site energies ($\epsilon_i=\epsilon_\mu=\epsilon$), therefore having $\mathcal{W}^{11}=\epsilon\mathbbm{1}_3$ in Eq.~(\ref{eq_schur}). Likewise, the other blocks are obtained by considering equal hopping strengths $t_{\mu i}=t$ ($\mu\neq a_1$) between pairs of external sites and $t_{a_1 i}=t'$ between the central site $a_1$ and its neighbors (see Fig.~\ref{fig:setup}). Here, thanks to the symmetry of the setup, we have again $\kappa_{12}=\kappa_{23}=\kappa_{31}=\kappa_{\text{cw}}$ and $\kappa_{13}=\kappa_{32}=\kappa_{21}=\kappa_{\overline{\text{cw}}}$ as for the basic three-sites ring. Moreover, $\kappa_{\overline{\text{cw}}}$ is obtained from $\kappa_{\text{cw}}$ by simply reversing the magnetic flux; therefore we can just focus on one of them. In Fig.~\ref{fig:kappa-7ring} we show the heat  conductance $\kappa_{\text{cw}}$ as a function of the magnetic flux, for $t=0.5\Gamma$ and various parameters. For a disconnected center site, $t'=0$, the plot is $2\pi$-periodic, as for the simplest three-sites model, because in this case any allowed path beginning and ending at the same terminal encloses the whole flux too. However, as soon as $t'\neq 0$, this is no longer true and different paths are available, thus changing the periodicity.  For this geometry, it is $12\pi$ because the minimal flux enclosed by a path starting and ending at the same terminal is $\alpha/6$ instead of $\alpha$. 
The other important aspect emerging from the plots is that the behavior of the heat conductance is quite sensitive to the variations of both $t'$ and $\epsilon$. This is reflected as well by the circulation coefficient, see Fig.~\ref{fig:kappa-7ring}(c), which depends in a highly nontrivial way on the model parameters. Note however, that despite this strong parameter dependence, in several instances a very high (and even maximal) circulation coefficient $\mathcal{C}$ can be obtained.
At the same time, this behavior indicates  less robustness in the circulation coefficient against sample-to-sample variations of the model parameters, compared to what we have previously illustrated in Fig.~\ref{fig:C-3ring-random} for the simple three-sites model.

\begin{figure}[t]
\centering
\includegraphics[width=0.9\columnwidth]{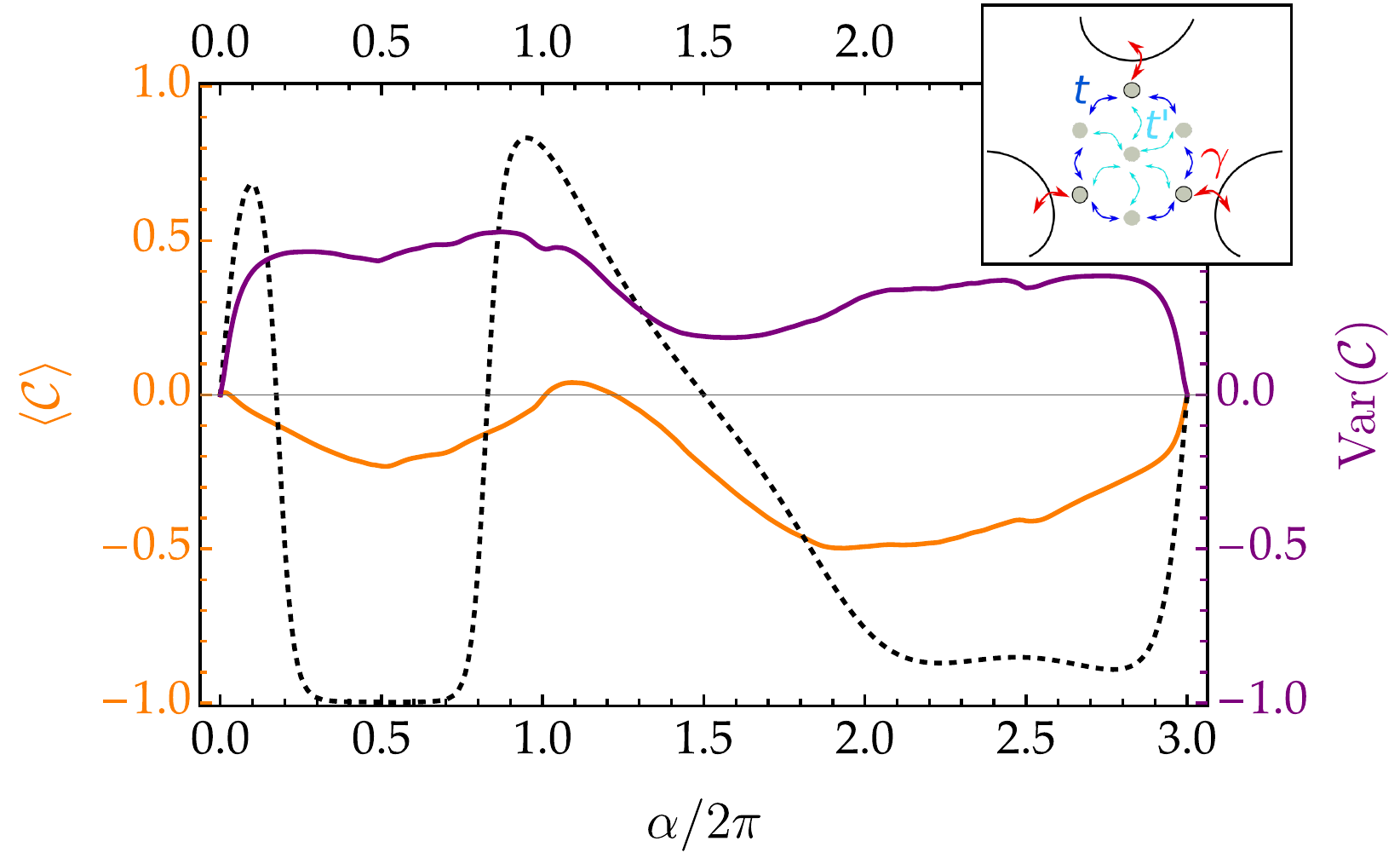}
\caption{Average circulation coefficient (solid orange line) and variance (solid purple line) for an ensemble of random hoppings and on-site energies, see main text. The dashed line denotes the results for the clean case with $\epsilon=0.3\Gamma$, $t=\Gamma$ and $t'=0.4\Gamma$. These plots refer to the setup shown in Fig.~\ref{fig:setup}(e).
\label{fig:C-7ring-random-2}}
\end{figure}

In Fig.~\ref{fig:C-7ring-random-2}, we show the averaged circulation coefficient over 2000 samples generated for a random choice of hoppings and on-site energies. As for the simple three-sites model, the parameters were allowed to vary independently from each other, meaning that the external and internal hopping amplitudes, as well as the on-site energies were not constrained to be equal among each other. In Fig.~\ref{fig:C-7ring-random-2}, we consider the average values $\braket{t_{\mu i}}=\Gamma$ ($\mu\neq a_1$), $t_{a_1 i}=0.4\Gamma$, $\braket{\epsilon_i}=0.3\Gamma$ and full widths $\delta(t_{\mu i})=0.6\Gamma$, $\delta(t_{a_1 i})=0.8\Gamma$, $\delta(\epsilon_i)=2\Gamma$.
The dashed black line shows the circulation coefficient, calculated for a fixed value of the parameters. This indeed shows that variations in the parameters have a much more pronounced impact on the magnitude of the circulation coefficient (compared to Fig.\ \ref{fig:C-3ring-random}). Also, parameter variations in the hexagonal model make the variance (purple line) to be of the same order of magnitude as the average value $\braket{\mathcal{C}}$. Note however, that the circulation effect is not fully suppressed but persists. 

\subsubsection{Square central structure}

\begin{figure}[t]
\centering
\includegraphics[width=\columnwidth]{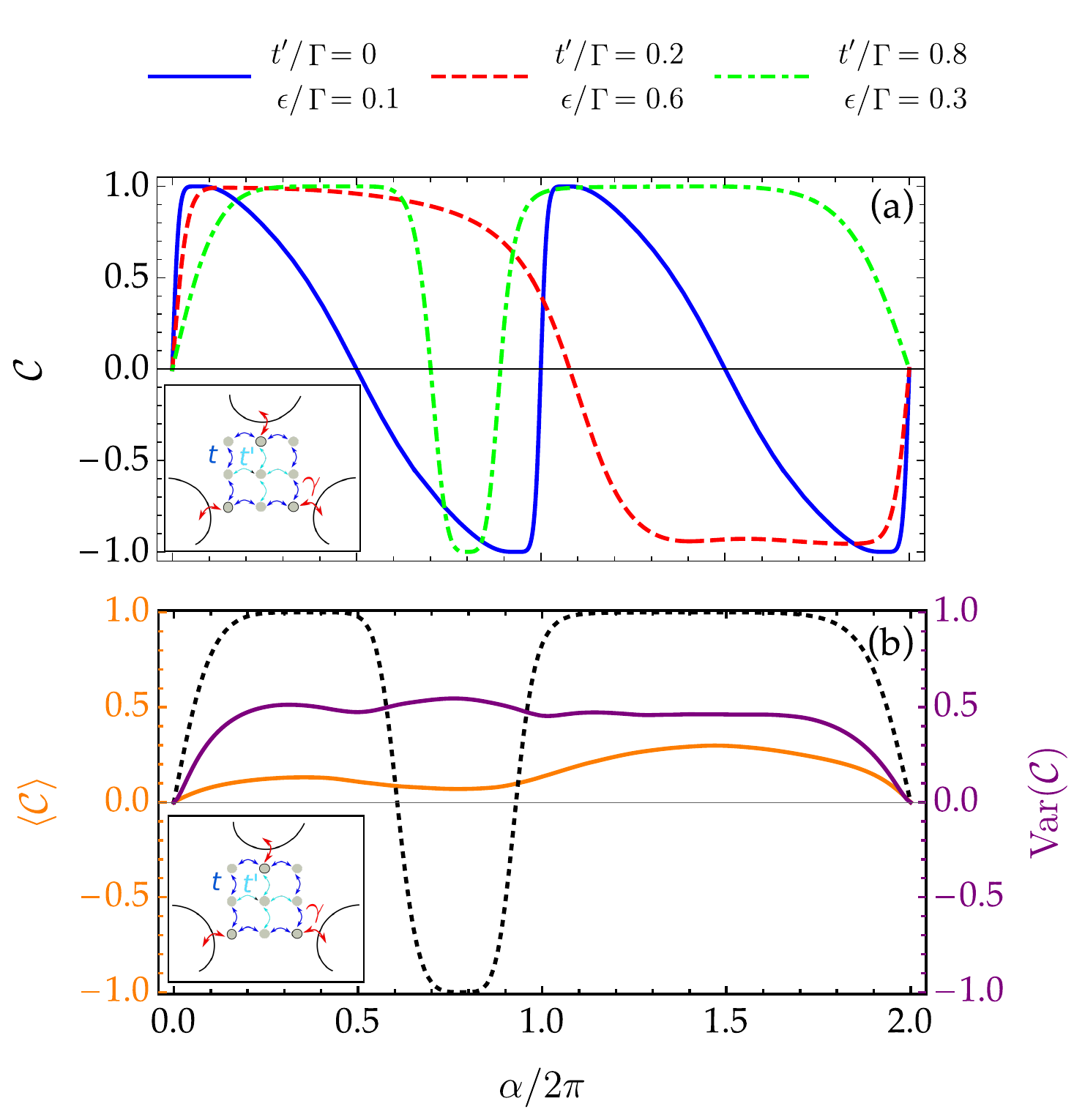}
\caption{Heat circulation with a square central structure. (a) Circulation coefficient for $t=0.5\Gamma$ and various combinations of $t'$ and $\epsilon$. (b) Ensemble-averaged $\braket{\mathcal{C}}$ (solid orange line), for a random variation of the parameters, compared with the corresponding result when no sample-to-sample variation is present (dashed black curve). The solid purple line shows the variance of $\mathcal{C}$.
\label{fig:C-9ring}}
\end{figure}

Let us now come to the analysis of the setup with a square structure, as sketched in Fig.\ \ref{fig:setup}(d). This configuration shows similar features as the hexagonal one, which we have just illustrated. Only details are different, due to the different symmetry of the system. For instance, the periodicity of the heat conductances of the square structure is $8\pi$, because (as soon as the internal hopping $t'$ is non zero) the minimal flux that can be enclosed in a path starting and ending at the same terminal is $\alpha/4$. Moreover, the position of the terminals is now such that not all heat  conductances are equal to each other: $\kappa_{23}=\kappa_{31}\neq\kappa_{12}$ (and similarly for the counterclockwise direction). However, apart from these specific differences, the qualitative behavior is the same: the system is quite sensitive to a variation of the model parameters, which results in a considerable reduction of the circulation coefficient when sample-to-sample variations are introduced. This is shown in Fig.~\ref{fig:C-9ring}, where we observe a stronger drop in the circulation coefficient, even compared to the one of the hexagonal structure in Fig.\ \ref{fig:C-7ring-random-2}. This can be attributed to the increased asymmetry of  the device with respect to the three contacts.

\subsubsection{Three-terminal chaotic cavity}\label{sec:chaotic}

Given the fact that, as we have seen, the performance of the system deteriorates considerably when increasing the complexity of the minimal model only slightly, it is an important question to understand how much circulation can be expected in more realistic models for extended scattering regions. To answer to this question, we study the case of an extended central system (cavity), e.g., a large quantum dot. The dynamics in such a system (with irregular boundaries) is chaotic, see Ref.~\cite{alhassid2000} for a review. Of course, this example is quite far apart from the initial simple ring-shaped system considered in Sec.~\ref{sec:N_model}. Nonetheless, we will see that heat circulation is still possible under certain conditions.

We start by investigating the system in the case where only a single mode ($M=1$) of each lead is coupled to the central cavity. Such a system can be realized by having a quantum point contact, tuned to the first conductance step, in between the cavity and each of the leads.
A successful method to address chaotic and disordered systems is random matrix theory~\cite{beenakker-review}. According to this approach, the scattering matrix $\mathcal{S}$ for a system with broken time-reversal symmetry is a random matrix distributed in the Circular Unitary Ensemble (CUE). However, random matrices from the CUE do not carry information on the magnetic field. An extension of the CUE has therefore been  developed to include the intensity of the magnetic field $B$ as a parameter, resulting in the formula \cite{brouwer96}
\begin{equation}
    \mathcal{S}(B)=U^{11}+U^{12}\left[\mathbbm{1}_\mathcal{N}-R(B)U^{22}\right]^{-1}R(B)U^{21}\,.
    \label{eq:random-S}
\end{equation}
Here, $U^{ij}$ are the four blocks of a $(3+\mathcal{N})\times(3+\mathcal{N})$ random matrix $U$ distributed according to the Circular Orthogonal Ensemble (unitary and symmetric matrices). Furthermore, we have $R(B)=\exp(B Q)$, $Q$ being an arbitrary real and antisymmetric matrix. 
As long as $\mathcal{N}\gg 1$, the detailed choice of the matrix $Q$ has been shown to be irrelevant and the result only depends on the parameter $\text{Tr}(Q
^2)$, related to the Thouless energy $E_\text{th}$ and the mean level spacing $\delta$ in the cavity \cite{brouwer96}. Therefore, it is convenient to use for $R$ the parametrization $R(x)=\exp(x D)$ \cite{meyer17}, $D$ being an antisymmetric matrix with $\text{Tr}(D^2)=-1$, while $x$ is a dimensionless quantity related to the magnetic flux piercing the cavity as $x\propto \alpha\sqrt{E_\text{Th}/\delta}$. The exact proportionality coefficient is a numerical factor of order $1$ that depends on the precise shape of the cavity \cite{brouwer96}. With this parametrization, the distribution of the scattering matrix $\mathcal{S}
$ interpolates between the Circular Orthogonal Ensemble (COE) at $x=0$ (time-reversal symmetric system) and the CUE at $x\gg1$ (when time reversal symmetry is fully broken). In contrast to the previous models, the scattering matrix obtained from Eq.\ \eqref{eq:random-S} is no longer periodic in $\alpha$ as the magnetic flux enclosed in the path between any two terminals of the system can assume arbitrary values.

\begin{figure}[t]
\centering
\includegraphics[width=\columnwidth]{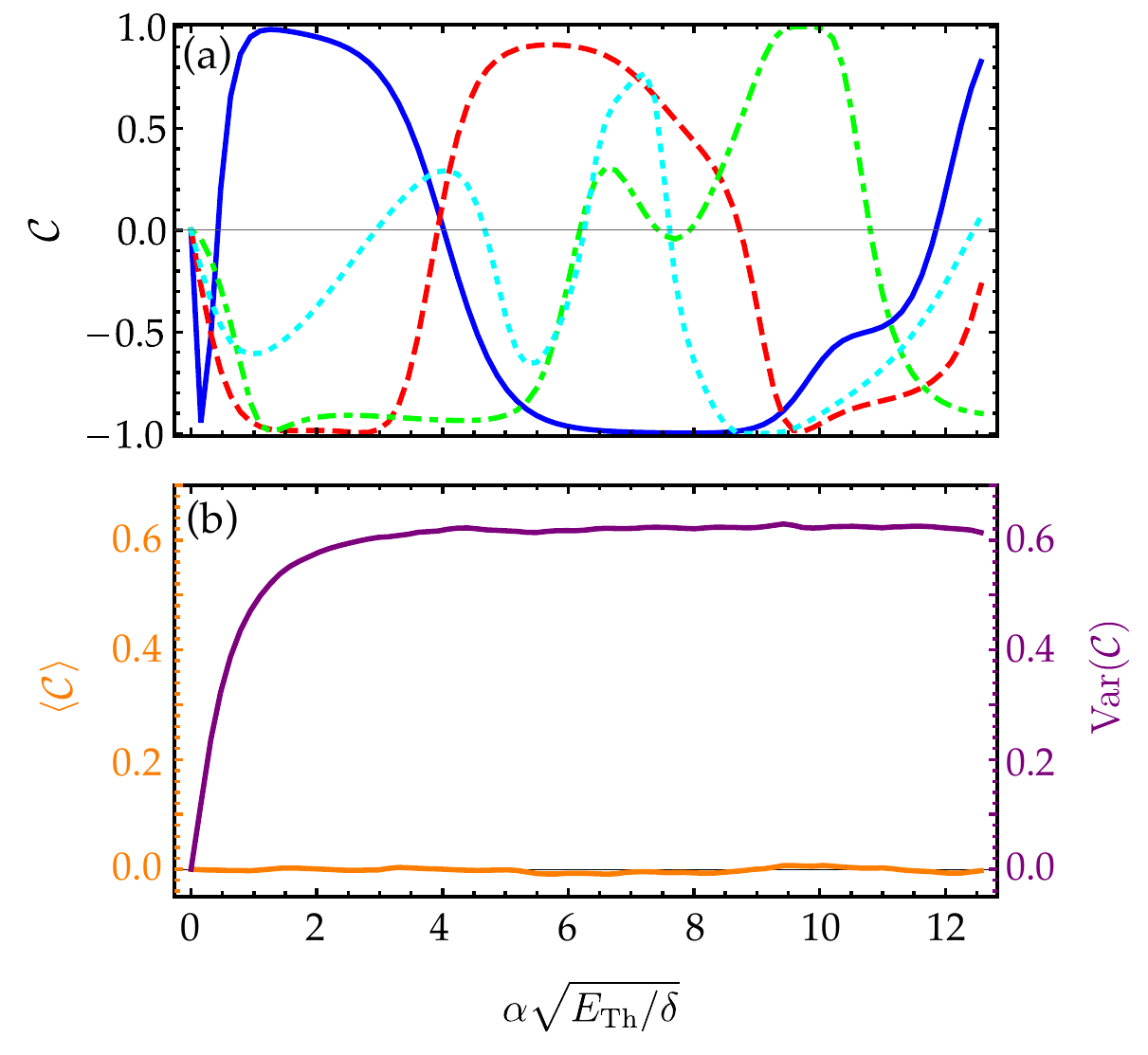}
\caption{(a) Circulation coefficient $\mathcal{C}$ for some realizations of the random scattering matrix \eqref{eq:random-S} describing the sample-to-sample variation of a chaotic cavity. (b) Averaged circulation coefficient (solid orange line) and its variance (solid purple line) over $10000$ random samples, generated according to Eq.\ \eqref{eq:random-S}, with $\mathcal{N}=40$.
\label{fig:C-chaotic}}
\end{figure}

\begin{figure}[t]
\centering
\includegraphics[width=\columnwidth]{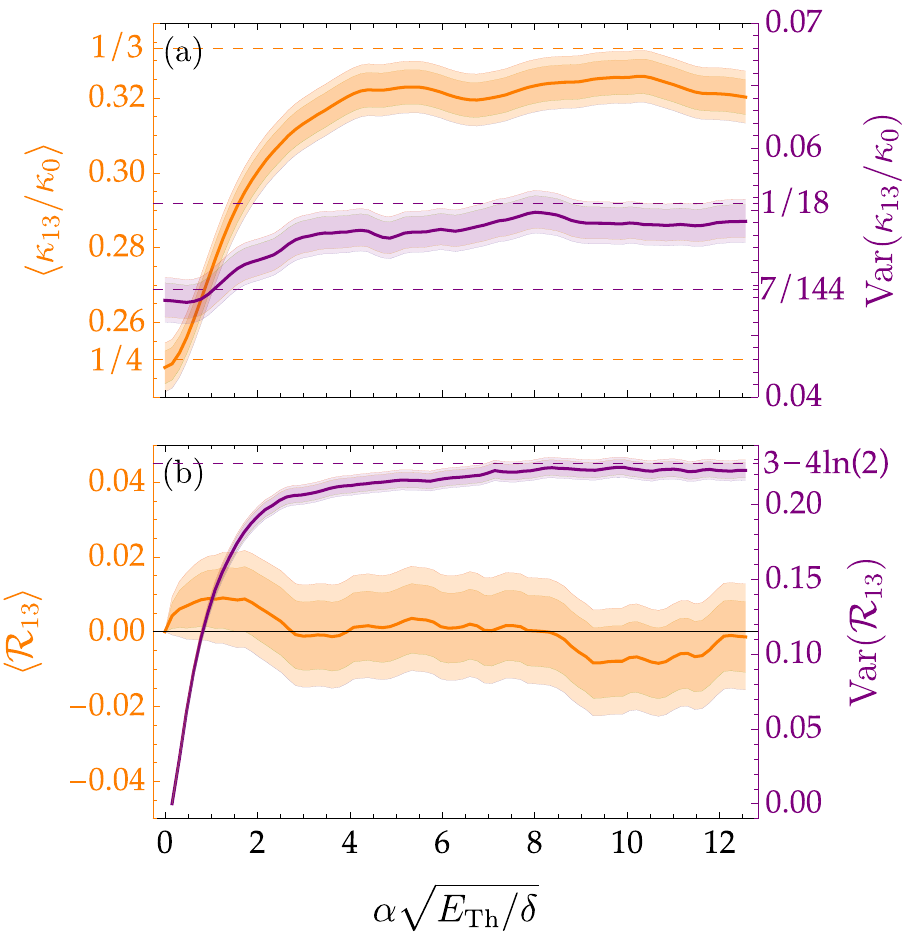}
\caption{(a) Averaged heat conductance $\Braket{\kappa_{13}}$ (orange) and its variance (purple). The dashed lines highlight the analytical predictions from Eqs.~\eqref{eq:k-CUE} and \eqref{eq:k-COE}. (b) Averaged rectification coefficient $\Braket{\mathcal{R}_{13}}$ (orange) and its variance (purple). The dashed line shows the analytical result from Eq.~\eqref{eq:variance-R}. In both panels the averages are taken over $10000$ random samples, generated according to Eq.~\eqref{eq:random-S}, with $\mathcal{N}=40$. Finally, the shaded bands correspond to 95\% (darker) and 99\% (lighter) confidence intervals.
\label{fig:k-R-chaotic}}
\end{figure}

The trend indicated in the two setups with extra lattice sites is continued in the chaotic system: in the same way as the variations of the model parameters produced quite different results for both the heat conductance and the circulation coefficient, here any different realization of the random scattering matrix \eqref{eq:random-S} results in a completely different outcome for the same quantities. Some examples illustrating this behavior are shown in Fig.\ \ref{fig:C-chaotic}(a). As a result, after averaging over a large number of random samples, the circulation coefficient is suppressed to zero. This is shown in Fig.\ \ref{fig:C-chaotic}(b), where the ensemble-averaged circulation coefficient $\mathcal{C}$ is plotted as a function of $x$. The average is performed over $10000$ samples, generated according to Eq.\ \eqref{eq:random-S}, with $\mathcal{N}=40$. It should be emphasized, however, that the sample-to-sample variations of $\mathcal{C}$ are quite large, as shown by the variance (purple curve); this confirms that it is likely that a given random realization produces a good circulation coefficient for some fine-tuned values of the magnetic flux, which depend on the specific device. In particular, Fig.\ \ref{fig:C-chaotic}(a) shows some realizations where a close-to-optimal circulation coefficient is reached for several values of $x\propto\alpha\sqrt{E_\text{Th}/\delta}$.

Let us now focus on the evolution of $\text{Var}(\mathcal{C})$ as a function of $x$: we observe a transition from a vanishing variance at small $x$ towards a more or less constant value when $x$ increases. This evolution is the result of the progressive breaking of time reversal symmetry. Indeed, at zero magnetic flux, we have $\mathcal{C}=0$ for every realization of the scattering matrix, since Onsager's reciprocity guarantees that $\kappa_{ij}=\kappa_{ji}$. As a result, also the variance $\text{Var}(\mathcal{C})$ vanishes in this case. On the other hand when time-reversal symmetry is broken, it is possible to have $\kappa_{ij}\neq\kappa_{ji}$ and therefore the circulation coefficient varies. In order to have an independent check of the magnitude of $\text{Var}(\mathcal{C})$ at large $x$, it is useful to find analytical expressions in limiting cases. We discuss in the following how to achieve this, considering that at large $x$ the scattering matrix is distributed in the CUE.

With this goal in mind, we first investigate the sample-to-sample variations of the heat conductance and the rectification coefficient. They are reported in Fig.\ \ref{fig:k-R-chaotic}, showing $\Braket{\kappa_{13}/\kappa_0}$ and $\Braket{\mathcal{R}_{13}}$, together with their variances. The values of the ensemble-averaged heat  conductance can be calculated analytically for $x=0$ and $x\gg1$. Indeed the ensemble averaging amounts to an integration in the unitary group and, in the case of the heat  conductance, Eq.\ \eqref{eq:kappa-normal} shows that we have to integrate a polynomial function of the scattering matrix $\mathcal{S}$. It is known how to perform such integrations (see for instance \cite{brouwer96b}) and we get (for $i\neq j$)
\begin{equation}
\begin{split}
    \Braket{\frac{\kappa_{ij}}{\kappa_0}}_{x\gg1}&=\frac{1}{d}=\frac{1}{3}\,,\\
    \text{Var}\left(\frac{\kappa_{ij}}{\kappa_0}\right)_{x\gg1}&=\frac{d^2-1}{d^2(d+1)}=\frac{1}{18}\,,
\end{split}
\label{eq:k-CUE}
\end{equation}
where we used that the size of the scattering matrix $\mathcal{S}$ is in our case $d=3$. These values are in good agreement with the numerical average in Fig.\ \ref{fig:k-R-chaotic}(a) for $x$ large enough. In a similar way, it is possible to obtain analytical results when $x=0$ and the scattering matrix is distributed in the COE. In this case we find
\begin{equation}
\begin{split}
    \Braket{\frac{\kappa_{ij}}{\kappa_0}}_{x=0}&=\frac{1}{d+1}=\frac{1}{4}\,,\\
    \text{Var}\left(\frac{\kappa_{ij}}{\kappa_0}\right)_{x=0}&=\frac{d^2+d+2}{d(d+1)^2(d+3)}=\frac{7}{144}\,,
\end{split}
\label{eq:k-COE}
\end{equation}
again in agreement with Fig.\ \ref{fig:k-R-chaotic}(a).

Let us now consider the rectification coefficient and the circulation coefficient. At zero magnetic flux, we have already observed that $\mathcal{R}_{ij}=\mathcal{C}=0$ for every realization. We then consider the case of large $x$ and do the averaging over the unitary group. It is easy to conclude on a general basis that $\Braket{\mathcal{R}_{ij}}=\Braket{\mathcal{C}}=0$ thanks to the possibility of relabelling the indices in the definitions \eqref{eq:rect} and \eqref{eq:C} when computing the integration over the unitary group (see App.~\ref{app:eq20}). Concerning the variance, we take a specific parametrization of $U(3)$ in terms of trigonometric functions \cite{bronzan1988} in order to compute ensemble averages. For the rectification we obtain (see App.~\ref{app:eq20})
\begin{equation}
    \text{Var}(\mathcal{R}_{ij})=\int_0^1 dy\int_0^1 dz\left(\frac{y-z}{y+z}\right)^2=3-4\ln 2\,,
    \label{eq:variance-R}
\end{equation}
for any $i\neq j$. Notice that the numerical value found in Fig.~\ref{fig:k-R-chaotic}(b) at large $x$ perfectly matches with the above analytical result. Finally, we have checked that the outcome of the integration over $U(3)$ is consistent with the value of $\text{Var}(\mathcal{C})$ found in Fig.~\ref{fig:C-chaotic}(b), although we are not able to provide an analytic result for this quantity.

\subsubsection{Chaotic cavity with multi-mode leads}\label{sec_multi}
\begin{figure}[t]
\centering
\includegraphics[width=0.9\columnwidth]{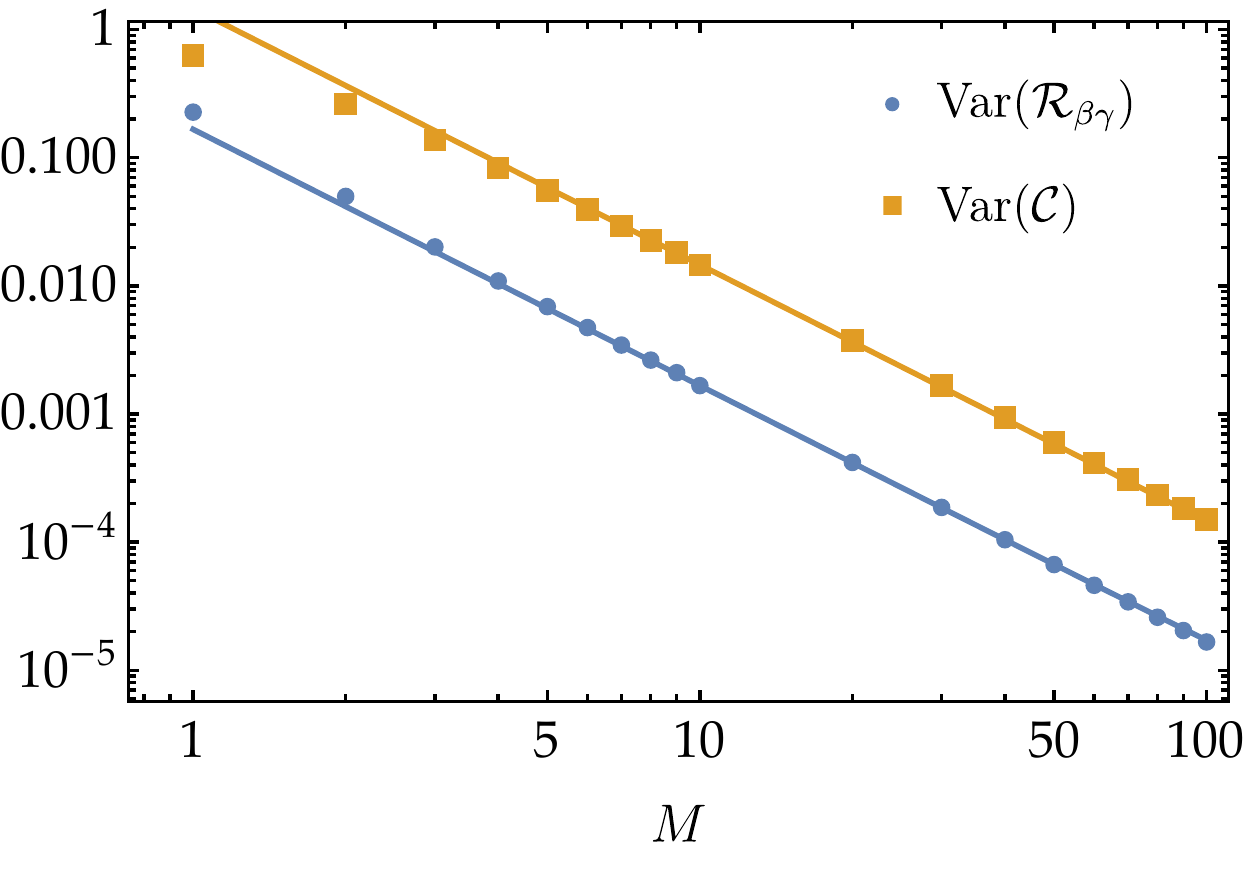}
\caption{Variance of the rectification (blue dots) and the circulation coefficient (orange squares) as a function of the channels $M$ in each of the three leads connected to the cavity. All points have been obtained by averaging over $10^5$ random samples. The lines are fits with the scaling law $\zeta M^{-2}$. The coefficients are $\zeta=1/6$ for $\mathcal{R}_{ij}$ and $\zeta=3/2$ for $\mathcal{C}$.
\label{fig:variance-multi}}
\end{figure}

So far, we have seen that when a single conduction channel connects the reservoirs to the chaotic cavity, the circulation is still highly efficient for some sample-specific values of the magnetic flux. Moreover, we have provided analytic results for the average and variance of heat conductances and rectification coefficients. It is natural to ask whether the circulation effect survives even when there are $M$ channels of each lead connected to the central cavity. As before, the average heat conductances and their variance can be obtained analytically. In the time-reversal-symmetric case $(x=0)$ we have
\begin{equation}
    \begin{split}
    \Braket{\frac{\kappa_{ij}}{\kappa_0}}_{x=0}&=\frac{M^2}{1+3M}\,,\\
    \text{Var}\left(\frac{\kappa_{ij}}{\kappa_0}\right)_{x=0}&=\frac{M(2+5M)}{9(1+3M)^2}\,,
    \end{split}
\end{equation}
whereas for broken time-reversal symmetry the result is
\begin{equation}
    \begin{split}
    \Braket{\frac{\kappa_{ij}}{\kappa_0}}_{x\gg1}&=\frac{M}{3}\,,\\
    \text{Var}\left(\frac{\kappa_{ij}}{\kappa_0}\right)_{x\gg1}&=\frac{4M^2}{9(9M^2-1)}\,.
    \end{split}
\end{equation}
Notice that for large $M$ there is no difference in the average heat conductance ($M/3$ in both cases), while a tiny difference persists in the variance ($4/81$ in the CUE and $5/81$ in the COE).
Next, we look at what happens to the rectification and the circulation coefficient by increasing $M$. We directly work in the limit $x\gg1$, so that the scattering matrix is distributed in the CUE and perform a numerical simulation by generating random matrices of increasing sizes. The result is shown in Fig.\ \ref{fig:variance-multi}, where we see that both $\text{Var}(\mathcal{R}_{ij})$ and $\text{Var}(\mathcal{C})$ decay as $M^{-2}$ for large $M$. This behavior signals that the different conducting channels are not independent of each other\footnote{The expected behavior in this case is $\text{Var}(\mathcal{C})\sim M^{-1}$ and similarly for $\text{Var}(\mathcal{R}_{ij})$.} and the coupling among them results in a faster decay of the variances of $\mathcal{R}_{ij}$ and $\mathcal{C}$. In addition, the decrease of the variances implies that sample-to-sample variations around the average become smaller and smaller by increasing $M$, meaning that, even for a single realization, the circulation coefficient is strongly suppressed if many channels are present. More precisely, starting from a fine-tuned value of the magnetic flux at which a given device has a good performance with one conducting channel, the circulation coefficient will decrease as $1/M$ by increasing the number of modes in the leads. This indicates that in order to maintain a good performance, the reservoirs have to be connected to the central scattering region via quantum point contacts, in such a way that at most a few conduction channels are open.

Finally, we investigate whether the $M^{-2}$ behavior observed in the numerical simulation, Fig.\ \ref{fig:variance-multi}, can also be understood analytically. In the following we show that indeed the decay of $\text{Var}(\mathcal{R}_{ij})$ and $\text{Var}(\mathcal{C})$ can be found exactly in the large-$M$ limit. Starting with the rectification, we can consider the approximation
\begin{equation}
    \text{Var}\left(\frac{X}{Y}\right)\approx \frac{\text{Var}(X)}{\Braket{Y}^2}+\frac{\Braket{X}^2}{\Braket{Y}^4}\text{Var}(Y)-\frac{2\Braket{X}}{\Braket{Y}^3}\text{Cov}(X,Y)
    \label{eq:approx-var}
\end{equation}
with $X=\kappa_{ij}-\kappa_{ji}$ and $Y=\kappa_{ij}+\kappa_{ji}$. This approximation is expected to work well when $X/Y$ does not depart too much from its average, which corresponds to the large-$M$ limit in our case. We clearly have $\Braket{X}=0$; moreover $\text{Cov}(X,Y)=0$ as can be verified via direct substitution. We have already calculated $\Braket{Y}=2\Braket{\kappa_{ij}}=2M/3$ and the only thing left is the evaluation of $\text{Var}(X)$, for which we obtain
\begin{equation}
    \text{Var}(\kappa_{ij}-\kappa_{ji})=\frac{2M^2}{3(9M^2-1)}\,.
\end{equation}
The approximation \eqref{eq:approx-var} then yields
\begin{equation}
    \text{Var}(\mathcal{R}_{ij})\approx\frac{3}{2(9M^2-1)}\underset{M\gg1}{\to}\frac{1}{6M^2}\,.
\end{equation}
This is exactly the behavior found with the best fit (blue line) in Fig.~\ref{fig:variance-multi}. Notice also that the above formula predicts $\text{Var}(\mathcal{R}_{ij})=3/16=0.1875$ for $M=1$, to be compared with the exact result $3-4\ln2=0.2274$ [see Eq.~\eqref{eq:variance-R}].
Concerning the circulation, we can apply the same method, with $X,Y=\kappa_{13}\kappa_{32}\kappa_{21}\mp\kappa_{12}\kappa_{23}\kappa_{31}$ and evaluate $\text{Var}(X)$ exactly for all $M$. However, the result is cumbersome (see Appendix \ref{app:M-exact}) and in the large-$M$ limit, we can adopt an alternative strategy, by considering a function
\begin{equation}
    g(x_1,\dots,x_6)=\frac{x_1x_2x_3-x_4x_5x_6}{x_1x_2x_3+x_4x_5x_6},
\end{equation}
where each $x_i$ is one heat conductance, according to the definition of $\mathcal{C}$. Its variance is estimated as
\begin{equation}
    \text{Var}(g)\approx\sum_{i=1}^6\left(\frac{\partial g}{\partial x_i}\right)^2\text{Var}(x_i)+2\sum_{i<j}\frac{\partial g}{\partial x_i}\frac{\partial g}{\partial x_j}\text{Cov}(x_i,x_j)
\end{equation}
with all derivatives being evaluated at $x_i=\Braket{x_i}$. The result of this calculation gives the estimate
\begin{equation}
    \text{Var}(\mathcal{C})\approx\frac{27}{2(9M^2-1)}\underset{M\gg1}{\to}\frac{3}{2M^2}\,,
    \label{eq:var-c-largeM}
\end{equation}
which again matches with the best fit in Fig.\ \ref{fig:variance-multi} and captures the $M^{-2}$ decay with the right prefactor.

\section{Conclusion}
\label{sec:conclusions}

In conclusion, we have presented a detailed analysis of three-terminal conductors with normal- or superconducting contacts acting as heat current circulators. We have shown that the presence of superconducting terminals as considered in a previous proposal~\cite{hwang18} is not an essential ingredient, even if they introduce further tunability on the device. Normalconducting systems have a similar (and often even improved) circulation coefficient. The essential requirement is the presence of a magnetic flux which breaks time-reversal symmetry.

Importantly, we have also investigated to what extent non ideal devices affect the circulation coefficient. In slightly modified setups compared to the proposal in~\cite{hwang18}, introducing the possibility of trajectories enclosing different amounts of magnetic flux, we found a much more important sensitivity on the system parameters. Therefore, the device is less robust with respect to sample-to-sample variations of these parameters, even though it is possible to fine-tune it to obtain high circulation coefficients. Finally, we addressed the extreme case of a chaotic scattering region and investigated  the statistics of the heat  conductances and the circulation performance. Here, while on average the circulation effect is completely suppressed, specific realizations still exhibit high performances, provided that the number of conducting channels is low.

An interesting issue to be still addressed is to understand the behavior of the heat current correlators (noise) in such multi-terminal devices and what information can be extracted from them. Despite being less studied with respect to its charge counterpart, heat current noise is an interesting topic for the community and is being more and more investigated
\cite{Krive2001Nov,Saito2007Oct,Saito2008Jul,Crepieux2014Dec,Crepieux2020Jul}. We leave this issue for future works.

\acknowledgements
We acknowledge helpful discussions with B. Sothmann, S.-Y. Hwang, T. L\"ofwander and G. Johansson. We acknowledge financial support from the Swedish VR (J.S.) and the Knut and Alice Wallenberg Foundation (M.A., F.Haj., and J.S.).

\appendix

\section{Derivation of the scattering matrix from a tight binding model}
\label{app:S-derivation}

In this Appendix we provide the details on the calculation of the scattering matrix $\mathcal{S}$ used in the main text. We work directly in the most general setup and obtain the result \eqref{eq:S-normal} as a particular case.

Let us consider $i=1,\dots, N$ semi-infinite chains with hopping amplitude $v$ between nearest neighbor sites. By labelling with $n_i$ the $n$-th site of the $i$-th chain we have the hopping Hamiltonian
\begin{equation}
    H_0=-v\sum_{n=-\infty}^{-1}\sum_{i=1}^N\left(\ket{n_i}\bra{n_i-1}+\ket{n_i-1}\bra{n_i}\right)
\end{equation}
The scattering region is formed by $N$ sites $(0_1,\dots 0_N)$, $0_i$ being connected to the $i$-th chain, and by
$\mathcal{M}$ additional sites that are not directly connected to the chains $(a_1,\dots,a_\mathcal{M})$. The central region is then made of $N+\mathcal{M}$ sites, coupled to each other via an $(N+\mathcal{M})\times(N+\mathcal{M})$ matrix $\mathcal{W}$. We write the corresponding Hamiltonian as
\begin{equation}
\begin{split}
    H_\text{s}&=\sum_{i,j=1}^{N}\mathcal{W}^{11}_{ij}\ket{0_i}\bra{0_j}+\sum_{\mu,\nu=a_1}^{a_\mathcal{M}}\mathcal{W}^{22}_{\mu\nu}\ket{\mu}\bra{\nu}\\
    &+\sum_{i=1}^N\sum_{\nu=a_1}^{a_\mathcal{M}}\mathcal{W}^{12}_{i\nu}\ket{0_i}\bra{\nu}+\sum_{\mu=a_1}^{a_\mathcal{M}}\sum_{j=1}^N\mathcal{W}^{21}_{\mu j}\ket{\mu}\bra{j}\,,
\end{split}
\end{equation}
where the block $\mathcal{W}^{11}$ ($\mathcal{W}^{22}$) of the matrix $\mathcal{W}$ describes the subspace of the sites $0_1,\dots,0_N$ ($a_1,\dots,a_\mathcal{M}$) only. Diagonal elements of these two blocks contain the onsite energies of the $N$ ($\mathcal{M}$) sites. The blocks $\mathcal{W}^{12}$ and $\mathcal{W}^{21}$ take cross couplings into account. Finally, the coupling between the scattering region and the leads is
\begin{equation}
    H_\text{c}=-\gamma\sum_{i=1}^{N}\left(\ket{0_i}\bra{-1_i}+\ket{-1_i}\bra{0_i}\right)\,.
\end{equation}

The free spectrum of the tight-binding chains is $E_q=-2v\cos q$. Now, in order to find the scattering matrix of the system, we consider an incoming wave from the chain $j$ and write the scattering state in the chain $i$ as (for $n\le -1$)
\begin{equation}
    \psi_{n_i}=\delta_{ij}e^{iqn}+\mathcal{S}_{ij}e^{-iqn}\,.
    \label{eq:app:scatt-state}
\end{equation}
Next, one has to solve the Schrödinger equation $\sum_\xi \braket{\zeta|H|\xi}\psi_\xi=E_q\psi_\zeta$, where $\xi$ and $\zeta$ can take values $n_i$ and $a_1,\dots,a_\mathcal{M}$. For $\zeta=-1_i$ we have
\begin{equation}
    -v\psi_{-2_i}-\gamma\psi_{0_i}=E_q\psi_{-1_i}\,.
    \label{eq:sch0}
\end{equation}
For $\zeta=0_i$ one finds
\begin{equation}
    -\gamma\psi_{-1_i}+\sum_{j=1}^{N}\mathcal{W}^{11}_{ij}\psi_{0_j}+\sum_{\nu=a_1}^{a_\mathcal{M}}\mathcal{W}^{12}_{i\nu}\psi_\nu=E_q\psi_{0_i}\,.
    \label{eq:app:sch1}
\end{equation}
Finally, for $\zeta=\mu=a_1,\dots,a_\mathcal{M}$,
\begin{equation}
    \sum_{j=1}^{N}W^{21}_{\mu j}\psi_{0_j}+\sum_{\nu=a_1}^{a_\mathcal{M}}W^{22}_{\mu\nu}\psi_\nu=E_q\psi_\mu\,.
    \label{eq:app:sch2}
\end{equation}
By eliminating $\psi_\mu$ and $\psi_{0_i}$ and recalling \eqref{eq:app:scatt-state} one eventually finds the matrix equation
\begin{widetext}
\begin{equation}
    -v(e^{-2iq}\mathbbm{1}_N+\mathcal{S}e^{2iq})-\gamma^2\left[\mathcal{W}^{11}+\mathcal{W}^{12}(E_q\mathbbm{1}_\mathcal{M}-\mathcal{W}^{22})^{-1}\mathcal{W}^{21}-E_q\mathbbm{1}_N\right]^{-1}(e^{-iq}\mathbbm{1}_N+\mathcal{S}e^{iq})=E_q(e^{-iq}\mathbbm{1}_N+\mathcal{S}e^{iq})\,,
\end{equation}
which is solved by (neglecting a global phase factor)
\begin{equation}
    \mathcal{S}=-\frac{(E_q+v e^{-iq})\mathbbm{1}_N+\gamma^2\left[\mathcal{W}^{11}+\mathcal{W}^{12}{(E_q\mathbbm{1}_\mathcal{M}-\mathcal{W}^{22})}^{-1}\mathcal{W}^{21}-E_q\mathbbm{1}_N\right]^{-1}}{(E_q+v e^{iq})\mathbbm{1}_N+\gamma^2\left[\mathcal{W}^{11}+\mathcal{W}^{12}{(E_q\mathbbm{1}_\mathcal{M}-\mathcal{W}^{22})}^{-1}\mathcal{W}^{21}-E_q\mathbbm{1}_N\right]^{-1}}\,,
\end{equation}
\end{widetext}
where the notation $A/B$ stands for $B^{-1}A$. Next, we linearize the spectrum $E_q\approx2v(q-\pi/2)$ (which amounts to approximate the density of states $\nu$ in the leads as a constant) and consider the wide-band limit, obtaining
\begin{equation}
    \mathcal{S}(E)=\left(\mathbbm{1}_N+\frac{i (W-E\mathbbm{1}_N)}{\Gamma}\right)^{-1}\left(\mathbbm{1}_N-\frac{i(W-E\mathbbm{1}_N)}{\Gamma}\right),
\end{equation}
where
$W=\mathcal{W}^{11}+\mathcal{W}^{12}{\left(E\mathbbm{1}_\mathcal{M}-\mathcal{W}^{22}\right)}^{-1}\mathcal{W}^{21}$ and $\Gamma=\gamma^2/v=\pi\gamma^2\nu$, $\nu$ being the density of states in the leads.
Finally, in the case where just the $N$ sites connected to the chains are present (and no additional ones), only the block $\mathcal{W}^{11}$ exists and the scattering matrix is found by simply letting $W=\mathcal{W}^{11}$. In this way, and also taking $N=3$, one recovers Eq.~\eqref{eq:S-normal}.

\section{Exact expressions for averages and variances}

\subsection{Derivation of Eq.\ \eqref{eq:variance-R}}
\label{app:eq20}
In this section we explain in more detail why the ensemble average of the rectification and the circulation yields $\Braket{\mathcal{R}_{ij}}=\Braket{\mathcal{C}}=0$ and we derive Eq.\ \eqref{eq:variance-R}. The ensemble average amounts to an integration over the unitary group: $\Braket{f(U)}=\int dV f(U)$, where $f$ is a generic function of a unitary matrix $U$ and $dV$ is the group measure. The latter is invariant if a unitary transformation is performed, meaning that $\int dV f(U)=\int dV f(U_0 U)$, where $U_0$ is unitary. In the case of $\mathcal{R}_{ij}$, $U$ is the scattering matrix and the function to be integrated is $f(U)=(|U_{ij}|^2-|U_{ji}|^2)/(|U_{ij}|^2+|U_{ji}|^2)=f_1(U)-f_2(U)$, where $f_1(U)=|U_{ij}|^2/(|U_{ij}|^2+|U_{ji}|^2)$ and $f_2(U)=|U_{ji}|^2/(|U_{ij}|^2+|U_{ji}|^2)$. It is clear that $f_1(U)=f_2(U_0U)$, $U_0$ being the unitary matrix which swaps rows and columns $(i,j)$ in the matrix $U$. Therefore \begin{equation}
\begin{split}
    \Braket{\mathcal{R}_{ij}}&=\int dV[f_1(U)-f_2(U)]\\
    &=\int dV f_1(U)-\int dV f_2(U_0U)=0\,.
\end{split}
\end{equation}
By the very same reasoning, one concludes that $\Braket{\mathcal{C}}=0$. 

We now evaluate $\text{Var}(\mathcal{R}_{ij})=\Braket{\mathcal{R}_{ij}^2}$ and derive Eq.\ \eqref{eq:variance-R}. First of all, following the same argument given above, one easily shows that $\text{Var}(\mathcal{R}_{ij})$ is the same for every $i\neq j$. Next, to get Eq.\ \eqref{eq:variance-R}, we consider a single channel for each lead and thus the integration is performed over the unitary group $U(3)$. We use the parametrization of Ref.\ \cite{bronzan1988}, according to which the group measure can be written as
\begin{equation}
    dV=-\frac{1}{32\pi^5}\prod_{i=1}^{5}d\phi_i\,d(\cos^4\theta_1)d(\cos^2\theta_2)d(\cos^2\theta_3)\,,
    \label{eq:app:group-meas}
\end{equation}
with $0\le\theta_1,\theta_2,\theta_3\le\pi/2$ and $0\le\phi_i\le 2\pi$. The parametrization of the elements $U_{ij}$ is given in Eq.\ (2.10) of Ref.\ \cite{bronzan1988}. As we said, $\text{Var}(\mathcal{R}_{ij})$ does not depend on $i$ and $j$ and then we can take $\mathcal{R}_{13}$ which is the most convenient for the calculation. By applying to the matrix $U$ the unitary transformation that exchanges the first two columns one gets
\begin{equation}
    \text{Var}(\mathcal{R}_{13})=\int dV\left[\frac{|U_{13}|^2-|U_{32}|^2}{|U_{13}|^2+|U_{32}|^2}\right]^2\,.
\end{equation}
The parametrization for the elements entering the last expression is \cite{bronzan1988} $U_{13}=\cos\theta_1\sin\theta_2 e^{i\phi_4}$ and $U_{32}=\cos\theta_1\sin\theta_3e^{i\phi_5}$. By using these relations in the previous formula, together with Eq.\ \eqref{eq:app:group-meas}, one finds
\begin{equation}
    \text{Var}(\mathcal{R}_{13})=\int_0^1\! d(\cos^2\!\theta_2)\int_0^1\! d(\cos^2\!\theta_3)\!\left[\frac{\sin^2\!\theta_2-\sin^2\!\theta_3}{\sin^2\!\theta_2+\sin^2\!\theta_3}\right]^2
\end{equation}
yielding Eq.\ \eqref{eq:variance-R} in the main text.

\subsection{Variance of the circulation coefficient}
\label{app:M-exact}
In this section we show a complementary way to arrive at the large-$M$ behavior $\text{Var}(\mathcal{C})\to 3/(2M^2)$ [see Eq.~\eqref{eq:var-c-largeM}]. Moreover, we also show exact expressions for the average and variances of the combinations of heat conductances entering the numerator of the circulation coefficient.

By using the diagrammatic method of Ref.\ \cite{brouwer96b} we find 
\begin{equation}
    \Braket{\kappa_{12}\kappa_{23}\kappa_{31}}=\Braket{\kappa_{13}\kappa_{32}\kappa_{21}}=\frac{M^5(9M^2-2)}{3(9M^2-1)(9M^2-4)}\,.
\end{equation}
Applying again the same technique, and using the tables in Ref.~\cite{stuart1980}, we evaluate $\mathcal{V}=\text{Var}(\kappa_{13}\kappa_{32}\kappa_{21}-\kappa_{12}\kappa_{23}\kappa_{31})$, finding $\mathcal{V}=11/420$ for $M=1$ and
\begin{widetext}
\begin{equation}
    \mathcal{V}=\frac{2M^4(6561M^{12}-34263M^{10}+50625M^8-14355M^6-6046M^4-2470M^2-84M+200)}{27(9M^2-1)^2(M^2-1)(9M^2-4)(9M^2-16)(9M^2-25)}
\end{equation}
for $M\ge 2$. According to Eq.~\eqref{eq:approx-var}, this gives an estimate for $\text{Var}(\mathcal{C})$ of $11/49$ when $M=1$ and
\begin{equation}
    \text{Var}(\mathcal{C})\approx\frac{(9M^2-4)(6561M^{12}-34263M^{10}+50625M^8-14355M^6-6046M^4-2470M^2-84M+200)}{6M^6(9M^2-2)^2(M^2-1)(9M^2-16)(9M^2-25)}
\end{equation}
\end{widetext}
for $M\ge 2$. As we can expect, this is a poor estimate for small $M$, but it captures exactly the large-$M$ behaviour $3M^{-2}/2$ and already for $M=8$ the error with respect to the true value is less than $5\%$.

\bibliography{Refs.bib}

\begin{thebibliography}{74}%
\makeatletter
\providecommand \@ifxundefined [1]{%
 \@ifx{#1\undefined}
}%
\providecommand \@ifnum [1]{%
 \ifnum #1\expandafter \@firstoftwo
 \else \expandafter \@secondoftwo
 \fi
}%
\providecommand \@ifx [1]{%
 \ifx #1\expandafter \@firstoftwo
 \else \expandafter \@secondoftwo
 \fi
}%
\providecommand \natexlab [1]{#1}%
\providecommand \enquote  [1]{``#1''}%
\providecommand \bibnamefont  [1]{#1}%
\providecommand \bibfnamefont [1]{#1}%
\providecommand \citenamefont [1]{#1}%
\providecommand \href@noop [0]{\@secondoftwo}%
\providecommand \href [0]{\begingroup \@sanitize@url \@href}%
\providecommand \@href[1]{\@@startlink{#1}\@@href}%
\providecommand \@@href[1]{\endgroup#1\@@endlink}%
\providecommand \@sanitize@url [0]{\catcode `\\12\catcode `\$12\catcode
  `\&12\catcode `\#12\catcode `\^12\catcode `\_12\catcode `\%12\relax}%
\providecommand \@@startlink[1]{}%
\providecommand \@@endlink[0]{}%
\providecommand \url  [0]{\begingroup\@sanitize@url \@url }%
\providecommand \@url [1]{\endgroup\@href {#1}{\urlprefix }}%
\providecommand \urlprefix  [0]{URL }%
\providecommand \Eprint [0]{\href }%
\providecommand \doibase [0]{http://dx.doi.org/}%
\providecommand \selectlanguage [0]{\@gobble}%
\providecommand \bibinfo  [0]{\@secondoftwo}%
\providecommand \bibfield  [0]{\@secondoftwo}%
\providecommand \translation [1]{[#1]}%
\providecommand \BibitemOpen [0]{}%
\providecommand \bibitemStop [0]{}%
\providecommand \bibitemNoStop [0]{.\EOS\space}%
\providecommand \EOS [0]{\spacefactor3000\relax}%
\providecommand \BibitemShut  [1]{\csname bibitem#1\endcsname}%
\let\auto@bib@innerbib\@empty
\bibitem [{\citenamefont {Li}\ \emph {et~al.}(2012)\citenamefont {Li},
  \citenamefont {Ren}, \citenamefont {Wang}, \citenamefont {Zhang},
  \citenamefont {H\"anggi},\ and\ \citenamefont {Li}}]{li2012}%
  \BibitemOpen
  \bibfield  {author} {\bibinfo {author} {\bibfnamefont {N.}~\bibnamefont
  {Li}}, \bibinfo {author} {\bibfnamefont {J.}~\bibnamefont {Ren}}, \bibinfo
  {author} {\bibfnamefont {L.}~\bibnamefont {Wang}}, \bibinfo {author}
  {\bibfnamefont {G.}~\bibnamefont {Zhang}}, \bibinfo {author} {\bibfnamefont
  {P.}~\bibnamefont {H\"anggi}}, \ and\ \bibinfo {author} {\bibfnamefont
  {B.}~\bibnamefont {Li}},\ }\bibfield  {title} {\enquote {\bibinfo {title}
  {{Colloquium: Phononics: Manipulating heat flow with electronic analogs and
  beyond}},}\ }\href {\doibase 10.1103/RevModPhys.84.1045} {\bibfield
  {journal} {\bibinfo  {journal} {Rev. Mod. Phys.}\ }\textbf {\bibinfo {volume}
  {84}},\ \bibinfo {pages} {1045--1066} (\bibinfo {year} {2012})}\BibitemShut
  {NoStop}%
\bibitem [{\citenamefont {Giazotto}\ \emph {et~al.}(2006)\citenamefont
  {Giazotto}, \citenamefont {Heikkil{\ifmmode\ddot{a}\else\"{a}\fi}},
  \citenamefont {Luukanen}, \citenamefont {Savin},\ and\ \citenamefont
  {Pekola}}]{Giazotto2006Mar}%
  \BibitemOpen
  \bibfield  {author} {\bibinfo {author} {\bibfnamefont {F.}~\bibnamefont
  {Giazotto}}, \bibinfo {author} {\bibfnamefont {T.~T.}\ \bibnamefont
  {Heikkil{\ifmmode\ddot{a}\else\"{a}\fi}}}, \bibinfo {author} {\bibfnamefont
  {A.}~\bibnamefont {Luukanen}}, \bibinfo {author} {\bibfnamefont {A.~M.}\
  \bibnamefont {Savin}}, \ and\ \bibinfo {author} {\bibfnamefont {J.~P.}\
  \bibnamefont {Pekola}},\ }\bibfield  {title} {\enquote {\bibinfo {title}
  {{Opportunities for mesoscopics in thermometry and refrigeration: Physics and
  applications}},}\ }\href {\doibase 10.1103/RevModPhys.78.217} {\bibfield
  {journal} {\bibinfo  {journal} {Rev. Mod. Phys.}\ }\textbf {\bibinfo {volume}
  {78}},\ \bibinfo {pages} {217--274} (\bibinfo {year} {2006})}\BibitemShut
  {NoStop}%
\bibitem [{\citenamefont {Pekola}\ \emph {et~al.}(2004)\citenamefont {Pekola},
  \citenamefont {Schoelkopf},\ and\ \citenamefont {Ullom}}]{pekola2004}%
  \BibitemOpen
  \bibfield  {author} {\bibinfo {author} {\bibfnamefont {J.}~\bibnamefont
  {Pekola}}, \bibinfo {author} {\bibfnamefont {R.}~\bibnamefont {Schoelkopf}},
  \ and\ \bibinfo {author} {\bibfnamefont {J.}~\bibnamefont {Ullom}},\
  }\bibfield  {title} {\enquote {\bibinfo {title} {{Cryogenics on a chip}},}\
  }\href {\doibase 10.1063/1.1768673} {\bibfield  {journal} {\bibinfo
  {journal} {Physics Today}\ }\textbf {\bibinfo {volume} {57}},\ \bibinfo
  {pages} {41} (\bibinfo {year} {2004})}\BibitemShut {NoStop}%
\bibitem [{\citenamefont {Paolucci}\ \emph {et~al.}(2018)\citenamefont
  {Paolucci}, \citenamefont {Marchegiani}, \citenamefont {Strambini},\ and\
  \citenamefont {Giazotto}}]{paolucci2018}%
  \BibitemOpen
  \bibfield  {author} {\bibinfo {author} {\bibfnamefont {F.}~\bibnamefont
  {Paolucci}}, \bibinfo {author} {\bibfnamefont {G.}~\bibnamefont
  {Marchegiani}}, \bibinfo {author} {\bibfnamefont {E.}~\bibnamefont
  {Strambini}}, \ and\ \bibinfo {author} {\bibfnamefont {F.}~\bibnamefont
  {Giazotto}},\ }\bibfield  {title} {\enquote {\bibinfo {title} {{Phase-Tunable
  Thermal Logic: Computation with Heat}},}\ }\href {\doibase
  10.1103/PhysRevApplied.10.024003} {\bibfield  {journal} {\bibinfo  {journal}
  {Phys. Rev. Applied}\ }\textbf {\bibinfo {volume} {10}},\ \bibinfo {pages}
  {024003} (\bibinfo {year} {2018})}\BibitemShut {NoStop}%
\bibitem [{\citenamefont {Guarcello}\ \emph {et~al.}(2018)\citenamefont
  {Guarcello}, \citenamefont {Solinas}, \citenamefont {Braggio}, \citenamefont
  {Di~Ventra},\ and\ \citenamefont {Giazotto}}]{guarcello2018}%
  \BibitemOpen
  \bibfield  {author} {\bibinfo {author} {\bibfnamefont {C.}~\bibnamefont
  {Guarcello}}, \bibinfo {author} {\bibfnamefont {P.}~\bibnamefont {Solinas}},
  \bibinfo {author} {\bibfnamefont {A.}~\bibnamefont {Braggio}}, \bibinfo
  {author} {\bibfnamefont {M.}~\bibnamefont {Di~Ventra}}, \ and\ \bibinfo
  {author} {\bibfnamefont {F.}~\bibnamefont {Giazotto}},\ }\bibfield  {title}
  {\enquote {\bibinfo {title} {{Josephson Thermal Memory}},}\ }\href {\doibase
  10.1103/PhysRevApplied.9.014021} {\bibfield  {journal} {\bibinfo  {journal}
  {Phys. Rev. Applied}\ }\textbf {\bibinfo {volume} {9}},\ \bibinfo {pages}
  {014021} (\bibinfo {year} {2018})}\BibitemShut {NoStop}%
\bibitem [{\citenamefont {Mart{\'i}nez-P{\'e}rez}\ \emph
  {et~al.}(2014)\citenamefont {Mart{\'i}nez-P{\'e}rez}, \citenamefont
  {Solinas},\ and\ \citenamefont {Giazotto}}]{m_perez2014}%
  \BibitemOpen
  \bibfield  {author} {\bibinfo {author} {\bibfnamefont {M.~J.}\ \bibnamefont
  {Mart{\'i}nez-P{\'e}rez}}, \bibinfo {author} {\bibfnamefont {P.}~\bibnamefont
  {Solinas}}, \ and\ \bibinfo {author} {\bibfnamefont {F.}~\bibnamefont
  {Giazotto}},\ }\bibfield  {title} {\enquote {\bibinfo {title} {{Coherent
  Caloritronics in Josephson-Based Nanocircuits}},}\ }\href {\doibase
  10.1007/s10909-014-1132-6} {\bibfield  {journal} {\bibinfo  {journal}
  {Journal of Low Temperature Physics}\ }\textbf {\bibinfo {volume} {175}},\
  \bibinfo {pages} {813--837} (\bibinfo {year} {2014})}\BibitemShut {NoStop}%
\bibitem [{\citenamefont {Fornieri}\ and\ \citenamefont
  {Giazotto}(2017)}]{Fornieri2017Oct}%
  \BibitemOpen
  \bibfield  {author} {\bibinfo {author} {\bibfnamefont {A.}~\bibnamefont
  {Fornieri}}\ and\ \bibinfo {author} {\bibfnamefont {F.}~\bibnamefont
  {Giazotto}},\ }\bibfield  {title} {\enquote {\bibinfo {title} {{Towards
  phase-coherent caloritronics in superconducting circuits}},}\ }\href
  {\doibase 10.1038/nnano.2017.204} {\bibfield  {journal} {\bibinfo  {journal}
  {Nat. Nanotechnol.}\ }\textbf {\bibinfo {volume} {12}},\ \bibinfo {pages}
  {944--952} (\bibinfo {year} {2017})}\BibitemShut {NoStop}%
\bibitem [{\citenamefont {Hwang}\ and\ \citenamefont
  {Sothmann}(2020)}]{Hwang2020-review}%
  \BibitemOpen
  \bibfield  {author} {\bibinfo {author} {\bibfnamefont {S.-Y.}\ \bibnamefont
  {Hwang}}\ and\ \bibinfo {author} {\bibfnamefont {B.}~\bibnamefont
  {Sothmann}},\ }\bibfield  {title} {\enquote {\bibinfo {title}
  {{Phase-coherent caloritronics with ordinary and topological Josephson
  junctions}},}\ }\href {https://doi.org/10.1140/epjst/e2019-900094-y}
  {\bibfield  {journal} {\bibinfo  {journal} {{Eur. Phys. J. Spec. Top.}}\
  }\textbf {\bibinfo {volume} {229}},\ \bibinfo {pages} {683--705} (\bibinfo
  {year} {2020})}\BibitemShut {NoStop}%
\bibitem [{\citenamefont {Terraneo}\ \emph {et~al.}(2002)\citenamefont
  {Terraneo}, \citenamefont {Peyrard},\ and\ \citenamefont
  {Casati}}]{terraneo2002}%
  \BibitemOpen
  \bibfield  {author} {\bibinfo {author} {\bibfnamefont {M.}~\bibnamefont
  {Terraneo}}, \bibinfo {author} {\bibfnamefont {M.}~\bibnamefont {Peyrard}}, \
  and\ \bibinfo {author} {\bibfnamefont {G.}~\bibnamefont {Casati}},\
  }\bibfield  {title} {\enquote {\bibinfo {title} {{Controlling the Energy Flow
  in Nonlinear Lattices: A Model for a Thermal Rectifier}},}\ }\href {\doibase
  10.1103/PhysRevLett.88.094302} {\bibfield  {journal} {\bibinfo  {journal}
  {Phys. Rev. Lett.}\ }\textbf {\bibinfo {volume} {88}},\ \bibinfo {pages}
  {094302} (\bibinfo {year} {2002})}\BibitemShut {NoStop}%
\bibitem [{\citenamefont {Li}\ \emph {et~al.}(2004)\citenamefont {Li},
  \citenamefont {Wang},\ and\ \citenamefont {Casati}}]{li2004}%
  \BibitemOpen
  \bibfield  {author} {\bibinfo {author} {\bibfnamefont {B.}~\bibnamefont
  {Li}}, \bibinfo {author} {\bibfnamefont {L.}~\bibnamefont {Wang}}, \ and\
  \bibinfo {author} {\bibfnamefont {G.}~\bibnamefont {Casati}},\ }\bibfield
  {title} {\enquote {\bibinfo {title} {{Thermal Diode: Rectification of Heat
  Flux}},}\ }\href {\doibase 10.1103/PhysRevLett.93.184301} {\bibfield
  {journal} {\bibinfo  {journal} {Phys. Rev. Lett.}\ }\textbf {\bibinfo
  {volume} {93}},\ \bibinfo {pages} {184301} (\bibinfo {year}
  {2004})}\BibitemShut {NoStop}%
\bibitem [{\citenamefont {Segal}\ and\ \citenamefont
  {Nitzan}(2005)}]{segal2005}%
  \BibitemOpen
  \bibfield  {author} {\bibinfo {author} {\bibfnamefont {D.}~\bibnamefont
  {Segal}}\ and\ \bibinfo {author} {\bibfnamefont {A.}~\bibnamefont {Nitzan}},\
  }\bibfield  {title} {\enquote {\bibinfo {title} {{Spin-Boson Thermal
  Rectifier}},}\ }\href {\doibase 10.1103/PhysRevLett.94.034301} {\bibfield
  {journal} {\bibinfo  {journal} {Phys. Rev. Lett.}\ }\textbf {\bibinfo
  {volume} {94}},\ \bibinfo {pages} {034301} (\bibinfo {year}
  {2005})}\BibitemShut {NoStop}%
\bibitem [{\citenamefont {Eckmann}\ and\ \citenamefont
  {Mej\'{\i}a-Monasterio}(2006)}]{eckmann2006}%
  \BibitemOpen
  \bibfield  {author} {\bibinfo {author} {\bibfnamefont {J.-P.}\ \bibnamefont
  {Eckmann}}\ and\ \bibinfo {author} {\bibfnamefont {C.}~\bibnamefont
  {Mej\'{\i}a-Monasterio}},\ }\bibfield  {title} {\enquote {\bibinfo {title}
  {{Thermal Rectification in Billiardlike Systems}},}\ }\href {\doibase
  10.1103/PhysRevLett.97.094301} {\bibfield  {journal} {\bibinfo  {journal}
  {Phys. Rev. Lett.}\ }\textbf {\bibinfo {volume} {97}},\ \bibinfo {pages}
  {094301} (\bibinfo {year} {2006})}\BibitemShut {NoStop}%
\bibitem [{\citenamefont {Zeng}\ and\ \citenamefont {Wang}(2008)}]{zeng2008}%
  \BibitemOpen
  \bibfield  {author} {\bibinfo {author} {\bibfnamefont {N.}~\bibnamefont
  {Zeng}}\ and\ \bibinfo {author} {\bibfnamefont {J.-S.}\ \bibnamefont
  {Wang}},\ }\bibfield  {title} {\enquote {\bibinfo {title} {{Mechanisms
  causing thermal rectification: The influence of phonon frequency}, asymmetry,
  and nonlinear interactions},}\ }\href {\doibase 10.1103/PhysRevB.78.024305}
  {\bibfield  {journal} {\bibinfo  {journal} {Phys. Rev. B}\ }\textbf {\bibinfo
  {volume} {78}},\ \bibinfo {pages} {024305} (\bibinfo {year}
  {2008})}\BibitemShut {NoStop}%
\bibitem [{\citenamefont {Ojanen}(2009)}]{ojanen2009}%
  \BibitemOpen
  \bibfield  {author} {\bibinfo {author} {\bibfnamefont {T.}~\bibnamefont
  {Ojanen}},\ }\bibfield  {title} {\enquote {\bibinfo {title} {{Selection-rule
  blockade and rectification in quantum heat transport}},}\ }\href {\doibase
  10.1103/PhysRevB.80.180301} {\bibfield  {journal} {\bibinfo  {journal} {Phys.
  Rev. B}\ }\textbf {\bibinfo {volume} {80}},\ \bibinfo {pages} {180301}
  (\bibinfo {year} {2009})}\BibitemShut {NoStop}%
\bibitem [{\citenamefont {Ruokola}\ \emph {et~al.}(2009)\citenamefont
  {Ruokola}, \citenamefont {Ojanen},\ and\ \citenamefont
  {Jauho}}]{ruokola2009}%
  \BibitemOpen
  \bibfield  {author} {\bibinfo {author} {\bibfnamefont {T.}~\bibnamefont
  {Ruokola}}, \bibinfo {author} {\bibfnamefont {T.}~\bibnamefont {Ojanen}}, \
  and\ \bibinfo {author} {\bibfnamefont {A.-P.}\ \bibnamefont {Jauho}},\
  }\bibfield  {title} {\enquote {\bibinfo {title} {{Thermal rectification in
  nonlinear quantum circuits}},}\ }\href {\doibase 10.1103/PhysRevB.79.144306}
  {\bibfield  {journal} {\bibinfo  {journal} {Phys. Rev. B}\ }\textbf {\bibinfo
  {volume} {79}},\ \bibinfo {pages} {144306} (\bibinfo {year}
  {2009})}\BibitemShut {NoStop}%
\bibitem [{\citenamefont {Wu}\ and\ \citenamefont {Segal}(2009)}]{wu2009}%
  \BibitemOpen
  \bibfield  {author} {\bibinfo {author} {\bibfnamefont {L.-A.}\ \bibnamefont
  {Wu}}\ and\ \bibinfo {author} {\bibfnamefont {D.}~\bibnamefont {Segal}},\
  }\bibfield  {title} {\enquote {\bibinfo {title} {{Sufficient Conditions for
  Thermal Rectification in Hybrid Quantum Structures}},}\ }\href {\doibase
  10.1103/PhysRevLett.102.095503} {\bibfield  {journal} {\bibinfo  {journal}
  {Phys. Rev. Lett.}\ }\textbf {\bibinfo {volume} {102}},\ \bibinfo {pages}
  {095503} (\bibinfo {year} {2009})}\BibitemShut {NoStop}%
\bibitem [{\citenamefont {Wu}\ \emph {et~al.}(2009)\citenamefont {Wu},
  \citenamefont {Yu},\ and\ \citenamefont {Segal}}]{wu2009b}%
  \BibitemOpen
  \bibfield  {author} {\bibinfo {author} {\bibfnamefont {L.-A.}\ \bibnamefont
  {Wu}}, \bibinfo {author} {\bibfnamefont {C.~X.}\ \bibnamefont {Yu}}, \ and\
  \bibinfo {author} {\bibfnamefont {D.}~\bibnamefont {Segal}},\ }\bibfield
  {title} {\enquote {\bibinfo {title} {{Nonlinear quantum heat transfer in
  hybrid structures: Sufficient conditions for thermal rectification}},}\
  }\href {\doibase 10.1103/PhysRevE.80.041103} {\bibfield  {journal} {\bibinfo
  {journal} {Phys. Rev. E}\ }\textbf {\bibinfo {volume} {80}},\ \bibinfo
  {pages} {041103} (\bibinfo {year} {2009})}\BibitemShut {NoStop}%
\bibitem [{\citenamefont {Kuo}\ and\ \citenamefont {Chang}(2010)}]{kuo2010}%
  \BibitemOpen
  \bibfield  {author} {\bibinfo {author} {\bibfnamefont {D.~M.-T.}\
  \bibnamefont {Kuo}}\ and\ \bibinfo {author} {\bibfnamefont {Y.-c.}\
  \bibnamefont {Chang}},\ }\bibfield  {title} {\enquote {\bibinfo {title}
  {Thermoelectric and thermal rectification properties of quantum dot
  junctions},}\ }\href {\doibase 10.1103/PhysRevB.81.205321} {\bibfield
  {journal} {\bibinfo  {journal} {Phys. Rev. B}\ }\textbf {\bibinfo {volume}
  {81}},\ \bibinfo {pages} {205321} (\bibinfo {year} {2010})}\BibitemShut
  {NoStop}%
\bibitem [{\citenamefont {Ruokola}\ and\ \citenamefont
  {Ojanen}(2011)}]{ruokola2011}%
  \BibitemOpen
  \bibfield  {author} {\bibinfo {author} {\bibfnamefont {T.}~\bibnamefont
  {Ruokola}}\ and\ \bibinfo {author} {\bibfnamefont {T.}~\bibnamefont
  {Ojanen}},\ }\bibfield  {title} {\enquote {\bibinfo {title} {{Single-electron
  heat diode: Asymmetric heat transport between electronic reservoirs through
  Coulomb islands}},}\ }\href {\doibase 10.1103/PhysRevB.83.241404} {\bibfield
  {journal} {\bibinfo  {journal} {Phys. Rev. B}\ }\textbf {\bibinfo {volume}
  {83}},\ \bibinfo {pages} {241404} (\bibinfo {year} {2011})}\BibitemShut
  {NoStop}%
\bibitem [{\citenamefont {Gunawardana}\ \emph {et~al.}(2012)\citenamefont
  {Gunawardana}, \citenamefont {Mullen}, \citenamefont {Hu}, \citenamefont
  {Chen},\ and\ \citenamefont {Ruan}}]{gunawardana2010}%
  \BibitemOpen
  \bibfield  {author} {\bibinfo {author} {\bibfnamefont {K.~G. S.~H.}\
  \bibnamefont {Gunawardana}}, \bibinfo {author} {\bibfnamefont
  {K.}~\bibnamefont {Mullen}}, \bibinfo {author} {\bibfnamefont
  {J.}~\bibnamefont {Hu}}, \bibinfo {author} {\bibfnamefont {Y.~P.}\
  \bibnamefont {Chen}}, \ and\ \bibinfo {author} {\bibfnamefont
  {X.}~\bibnamefont {Ruan}},\ }\bibfield  {title} {\enquote {\bibinfo {title}
  {Tunable thermal transport and thermal rectification in strained graphene
  nanoribbons},}\ }\href {\doibase 10.1103/PhysRevB.85.245417} {\bibfield
  {journal} {\bibinfo  {journal} {Phys. Rev. B}\ }\textbf {\bibinfo {volume}
  {85}},\ \bibinfo {pages} {245417} (\bibinfo {year} {2012})}\BibitemShut
  {NoStop}%
\bibitem [{\citenamefont {Mart\'{i}nez-P\'{e}rez}\ and\ \citenamefont
  {Giazotto}(2013{\natexlab{a}})}]{martinez2013}%
  \BibitemOpen
  \bibfield  {author} {\bibinfo {author} {\bibfnamefont {M.~J.}\ \bibnamefont
  {Mart\'{i}nez-P\'{e}rez}}\ and\ \bibinfo {author} {\bibfnamefont
  {F.}~\bibnamefont {Giazotto}},\ }\bibfield  {title} {\enquote {\bibinfo
  {title} {{Efficient phase-tunable Josephson thermal rectifier}},}\ }\href
  {\doibase 10.1063/1.4804550} {\bibfield  {journal} {\bibinfo  {journal}
  {Applied Physics Letters}\ }\textbf {\bibinfo {volume} {102}},\ \bibinfo
  {pages} {182602} (\bibinfo {year} {2013}{\natexlab{a}})}\BibitemShut
  {NoStop}%
\bibitem [{\citenamefont {Fornieri}\ \emph {et~al.}(2014)\citenamefont
  {Fornieri}, \citenamefont {Mart\'{i}nez-P\'{e}rez},\ and\ \citenamefont
  {Giazotto}}]{fornieri2014}%
  \BibitemOpen
  \bibfield  {author} {\bibinfo {author} {\bibfnamefont {A.}~\bibnamefont
  {Fornieri}}, \bibinfo {author} {\bibfnamefont {M.~J.}\ \bibnamefont
  {Mart\'{i}nez-P\'{e}rez}}, \ and\ \bibinfo {author} {\bibfnamefont
  {F.}~\bibnamefont {Giazotto}},\ }\bibfield  {title} {\enquote {\bibinfo
  {title} {{A normal metal tunnel-junction heat diode}},}\ }\href {\doibase
  10.1063/1.4875917} {\bibfield  {journal} {\bibinfo  {journal} {Applied
  Physics Letters}\ }\textbf {\bibinfo {volume} {104}},\ \bibinfo {pages}
  {183108} (\bibinfo {year} {2014})}\BibitemShut {NoStop}%
\bibitem [{\citenamefont {Landi}\ \emph {et~al.}(2014)\citenamefont {Landi},
  \citenamefont {Novais}, \citenamefont {de~Oliveira},\ and\ \citenamefont
  {Karevski}}]{landi2014}%
  \BibitemOpen
  \bibfield  {author} {\bibinfo {author} {\bibfnamefont {G.~T.}\ \bibnamefont
  {Landi}}, \bibinfo {author} {\bibfnamefont {E.}~\bibnamefont {Novais}},
  \bibinfo {author} {\bibfnamefont {M.~J.}\ \bibnamefont {de~Oliveira}}, \ and\
  \bibinfo {author} {\bibfnamefont {D.}~\bibnamefont {Karevski}},\ }\bibfield
  {title} {\enquote {\bibinfo {title} {{Flux rectification in the quantum $XXZ$
  chain}},}\ }\href {\doibase 10.1103/PhysRevE.90.042142} {\bibfield  {journal}
  {\bibinfo  {journal} {Phys. Rev. E}\ }\textbf {\bibinfo {volume} {90}},\
  \bibinfo {pages} {042142} (\bibinfo {year} {2014})}\BibitemShut {NoStop}%
\bibitem [{\citenamefont {S{\'{a}}nchez}\ \emph {et~al.}(2015)\citenamefont
  {S{\'{a}}nchez}, \citenamefont {Sothmann},\ and\ \citenamefont
  {Jordan}}]{sanchez2015}%
  \BibitemOpen
  \bibfield  {author} {\bibinfo {author} {\bibfnamefont {R.}~\bibnamefont
  {S{\'{a}}nchez}}, \bibinfo {author} {\bibfnamefont {B.}~\bibnamefont
  {Sothmann}}, \ and\ \bibinfo {author} {\bibfnamefont {A.~N.}\ \bibnamefont
  {Jordan}},\ }\bibfield  {title} {\enquote {\bibinfo {title} {{Heat diode and
  engine based on quantum Hall edge states}},}\ }\href {\doibase
  10.1088/1367-2630/17/7/075006} {\bibfield  {journal} {\bibinfo  {journal}
  {New Journal of Physics}\ }\textbf {\bibinfo {volume} {17}},\ \bibinfo
  {pages} {075006} (\bibinfo {year} {2015})}\BibitemShut {NoStop}%
\bibitem [{\citenamefont {Jiang}\ \emph {et~al.}(2015)\citenamefont {Jiang},
  \citenamefont {Kulkarni}, \citenamefont {Segal},\ and\ \citenamefont
  {Imry}}]{jiang2015}%
  \BibitemOpen
  \bibfield  {author} {\bibinfo {author} {\bibfnamefont {J.-H.}\ \bibnamefont
  {Jiang}}, \bibinfo {author} {\bibfnamefont {M.}~\bibnamefont {Kulkarni}},
  \bibinfo {author} {\bibfnamefont {D.}~\bibnamefont {Segal}}, \ and\ \bibinfo
  {author} {\bibfnamefont {Y.}~\bibnamefont {Imry}},\ }\bibfield  {title}
  {\enquote {\bibinfo {title} {Phonon thermoelectric transistors and
  rectifiers},}\ }\href {\doibase 10.1103/PhysRevB.92.045309} {\bibfield
  {journal} {\bibinfo  {journal} {Phys. Rev. B}\ }\textbf {\bibinfo {volume}
  {92}},\ \bibinfo {pages} {045309} (\bibinfo {year} {2015})}\BibitemShut
  {NoStop}%
\bibitem [{\citenamefont {Joulain}\ \emph {et~al.}(2016)\citenamefont
  {Joulain}, \citenamefont {Drevillon}, \citenamefont {Ezzahri},\ and\
  \citenamefont {Ordonez-Miranda}}]{joulain2016}%
  \BibitemOpen
  \bibfield  {author} {\bibinfo {author} {\bibfnamefont {K.}~\bibnamefont
  {Joulain}}, \bibinfo {author} {\bibfnamefont {J.}~\bibnamefont {Drevillon}},
  \bibinfo {author} {\bibfnamefont {Y.}~\bibnamefont {Ezzahri}}, \ and\
  \bibinfo {author} {\bibfnamefont {J.}~\bibnamefont {Ordonez-Miranda}},\
  }\bibfield  {title} {\enquote {\bibinfo {title} {{Quantum Thermal
  Transistor}},}\ }\href {\doibase 10.1103/PhysRevLett.116.200601} {\bibfield
  {journal} {\bibinfo  {journal} {Phys. Rev. Lett.}\ }\textbf {\bibinfo
  {volume} {116}},\ \bibinfo {pages} {200601} (\bibinfo {year}
  {2016})}\BibitemShut {NoStop}%
\bibitem [{\citenamefont {Marcos-Vicioso}\ \emph {et~al.}(2018)\citenamefont
  {Marcos-Vicioso}, \citenamefont {L\'opez-Jurado}, \citenamefont
  {Ruiz-Garcia},\ and\ \citenamefont {S\'anchez}}]{vicioso2018}%
  \BibitemOpen
  \bibfield  {author} {\bibinfo {author} {\bibfnamefont {A.}~\bibnamefont
  {Marcos-Vicioso}}, \bibinfo {author} {\bibfnamefont {C.}~\bibnamefont
  {L\'opez-Jurado}}, \bibinfo {author} {\bibfnamefont {M.}~\bibnamefont
  {Ruiz-Garcia}}, \ and\ \bibinfo {author} {\bibfnamefont {R.}~\bibnamefont
  {S\'anchez}},\ }\bibfield  {title} {\enquote {\bibinfo {title} {{Thermal
  rectification with interacting electronic channels: Exploiting degeneracy,
  quantum superpositions, and interference}},}\ }\href {\doibase
  10.1103/PhysRevB.98.035414} {\bibfield  {journal} {\bibinfo  {journal} {Phys.
  Rev. B}\ }\textbf {\bibinfo {volume} {98}},\ \bibinfo {pages} {035414}
  (\bibinfo {year} {2018})}\BibitemShut {NoStop}%
\bibitem [{\citenamefont {Bours}\ \emph {et~al.}(2019)\citenamefont {Bours},
  \citenamefont {Sothmann}, \citenamefont {Carrega}, \citenamefont {Strambini},
  \citenamefont {Braggio}, \citenamefont {Hankiewicz}, \citenamefont
  {Molenkamp},\ and\ \citenamefont {Giazotto}}]{bours2019}%
  \BibitemOpen
  \bibfield  {author} {\bibinfo {author} {\bibfnamefont {L.}~\bibnamefont
  {Bours}}, \bibinfo {author} {\bibfnamefont {B.}~\bibnamefont {Sothmann}},
  \bibinfo {author} {\bibfnamefont {M.}~\bibnamefont {Carrega}}, \bibinfo
  {author} {\bibfnamefont {E.}~\bibnamefont {Strambini}}, \bibinfo {author}
  {\bibfnamefont {A.}~\bibnamefont {Braggio}}, \bibinfo {author} {\bibfnamefont
  {E.~M.}\ \bibnamefont {Hankiewicz}}, \bibinfo {author} {\bibfnamefont
  {L.~W.}\ \bibnamefont {Molenkamp}}, \ and\ \bibinfo {author} {\bibfnamefont
  {F.}~\bibnamefont {Giazotto}},\ }\bibfield  {title} {\enquote {\bibinfo
  {title} {{Phase-Tunable Thermal Rectification in the Topological SQUIPT}},}\
  }\href {\doibase 10.1103/PhysRevApplied.11.044073} {\bibfield  {journal}
  {\bibinfo  {journal} {Phys. Rev. Applied}\ }\textbf {\bibinfo {volume}
  {11}},\ \bibinfo {pages} {044073} (\bibinfo {year} {2019})}\BibitemShut
  {NoStop}%
\bibitem [{\citenamefont {Goury}\ and\ \citenamefont
  {S\'{a}nchez}(2019)}]{goury2019}%
  \BibitemOpen
  \bibfield  {author} {\bibinfo {author} {\bibfnamefont {D.}~\bibnamefont
  {Goury}}\ and\ \bibinfo {author} {\bibfnamefont {R.}~\bibnamefont
  {S\'{a}nchez}},\ }\bibfield  {title} {\enquote {\bibinfo {title} {Reversible
  thermal diode and energy harvester with a superconducting quantum
  interference single-electron transistor},}\ }\href {\doibase
  10.1063/1.5109100} {\bibfield  {journal} {\bibinfo  {journal} {Applied
  Physics Letters}\ }\textbf {\bibinfo {volume} {115}},\ \bibinfo {pages}
  {092601} (\bibinfo {year} {2019})}\BibitemShut {NoStop}%
\bibitem [{\citenamefont {Giazotto}\ and\ \citenamefont
  {Bergeret}(2020)}]{giazotto2020}%
  \BibitemOpen
  \bibfield  {author} {\bibinfo {author} {\bibfnamefont {F.}~\bibnamefont
  {Giazotto}}\ and\ \bibinfo {author} {\bibfnamefont {F.~S.}\ \bibnamefont
  {Bergeret}},\ }\bibfield  {title} {\enquote {\bibinfo {title} {Very large
  thermal rectification in ferromagnetic insulator-based superconducting tunnel
  junctions},}\ }\href {\doibase 10.1063/5.0010148} {\bibfield  {journal}
  {\bibinfo  {journal} {Applied Physics Letters}\ }\textbf {\bibinfo {volume}
  {116}},\ \bibinfo {pages} {192601} (\bibinfo {year} {2020})}\BibitemShut
  {NoStop}%
\bibitem [{\citenamefont {Chang}\ \emph {et~al.}(2006)\citenamefont {Chang},
  \citenamefont {Okawa}, \citenamefont {Majumdar},\ and\ \citenamefont
  {Zettl}}]{Chang2006}%
  \BibitemOpen
  \bibfield  {author} {\bibinfo {author} {\bibfnamefont {C.~W.}\ \bibnamefont
  {Chang}}, \bibinfo {author} {\bibfnamefont {D.}~\bibnamefont {Okawa}},
  \bibinfo {author} {\bibfnamefont {A.}~\bibnamefont {Majumdar}}, \ and\
  \bibinfo {author} {\bibfnamefont {A.}~\bibnamefont {Zettl}},\ }\bibfield
  {title} {\enquote {\bibinfo {title} {{Solid-State Thermal Rectifier}},}\
  }\href {\doibase 10.1126/science.1132898} {\bibfield  {journal} {\bibinfo
  {journal} {Science}\ }\textbf {\bibinfo {volume} {314}},\ \bibinfo {pages}
  {1121--1124} (\bibinfo {year} {2006})}\BibitemShut {NoStop}%
\bibitem [{\citenamefont {Schmotz}\ \emph {et~al.}(2011)\citenamefont
  {Schmotz}, \citenamefont {Maier}, \citenamefont {Scheer},\ and\ \citenamefont
  {Leiderer}}]{Schmotz2011}%
  \BibitemOpen
  \bibfield  {author} {\bibinfo {author} {\bibfnamefont {M.}~\bibnamefont
  {Schmotz}}, \bibinfo {author} {\bibfnamefont {J.}~\bibnamefont {Maier}},
  \bibinfo {author} {\bibfnamefont {E.}~\bibnamefont {Scheer}}, \ and\ \bibinfo
  {author} {\bibfnamefont {P.}~\bibnamefont {Leiderer}},\ }\bibfield  {title}
  {\enquote {\bibinfo {title} {{A thermal diode using phonon rectification}},}\
  }\href {\doibase 10.1088/1367-2630/13/11/113027} {\bibfield  {journal}
  {\bibinfo  {journal} {New Journal of Physics}\ }\textbf {\bibinfo {volume}
  {13}},\ \bibinfo {pages} {113027} (\bibinfo {year} {2011})}\BibitemShut
  {NoStop}%
\bibitem [{\citenamefont {Mart{\'i}nez-P{\'e}rez}\ \emph
  {et~al.}(2015)\citenamefont {Mart{\'i}nez-P{\'e}rez}, \citenamefont
  {Fornieri},\ and\ \citenamefont {Giazotto}}]{Martinez2015}%
  \BibitemOpen
  \bibfield  {author} {\bibinfo {author} {\bibfnamefont {M.~J.}\ \bibnamefont
  {Mart{\'i}nez-P{\'e}rez}}, \bibinfo {author} {\bibfnamefont {A.}~\bibnamefont
  {Fornieri}}, \ and\ \bibinfo {author} {\bibfnamefont {F.}~\bibnamefont
  {Giazotto}},\ }\bibfield  {title} {\enquote {\bibinfo {title} {{Rectification
  of electronic heat current by a hybrid thermal diode}},}\ }\href {\doibase
  10.1038/nnano.2015.11} {\bibfield  {journal} {\bibinfo  {journal} {Nat.
  Nanotechnol.}\ }\textbf {\bibinfo {volume} {10}},\ \bibinfo {pages} {303}
  (\bibinfo {year} {2015})}\BibitemShut {NoStop}%
\bibitem [{\citenamefont {Scheibner}\ \emph {et~al.}(2008)\citenamefont
  {Scheibner}, \citenamefont {König}, \citenamefont {Reuter}, \citenamefont
  {Wieck}, \citenamefont {Gould}, \citenamefont {Buhmann},\ and\ \citenamefont
  {Molenkamp}}]{Scheibner_2008}%
  \BibitemOpen
  \bibfield  {author} {\bibinfo {author} {\bibfnamefont {R.}~\bibnamefont
  {Scheibner}}, \bibinfo {author} {\bibfnamefont {M.}~\bibnamefont {König}},
  \bibinfo {author} {\bibfnamefont {D.}~\bibnamefont {Reuter}}, \bibinfo
  {author} {\bibfnamefont {A.~D.}\ \bibnamefont {Wieck}}, \bibinfo {author}
  {\bibfnamefont {C.}~\bibnamefont {Gould}}, \bibinfo {author} {\bibfnamefont
  {H.}~\bibnamefont {Buhmann}}, \ and\ \bibinfo {author} {\bibfnamefont
  {L.~W.}\ \bibnamefont {Molenkamp}},\ }\bibfield  {title} {\enquote {\bibinfo
  {title} {Quantum dot as thermal rectifier},}\ }\href {\doibase
  10.1088/1367-2630/10/8/083016} {\bibfield  {journal} {\bibinfo  {journal}
  {New Journal of Physics}\ }\textbf {\bibinfo {volume} {10}},\ \bibinfo
  {pages} {083016} (\bibinfo {year} {2008})}\BibitemShut {NoStop}%
\bibitem [{\citenamefont {Partanen}\ \emph {et~al.}(2018)\citenamefont
  {Partanen}, \citenamefont {Tan}, \citenamefont {Masuda}, \citenamefont
  {Govenius}, \citenamefont {Lake}, \citenamefont {Jenei}, \citenamefont
  {Gr{\"o}nberg}, \citenamefont {Hassel}, \citenamefont {Simbierowicz},
  \citenamefont {Vesterinen}, \citenamefont {Tuorila}, \citenamefont
  {Ala-Nissila},\ and\ \citenamefont {M{\"o}tt{\"o}nen}}]{Partanen2018}%
  \BibitemOpen
  \bibfield  {author} {\bibinfo {author} {\bibfnamefont {M.}~\bibnamefont
  {Partanen}}, \bibinfo {author} {\bibfnamefont {K.~Y.}\ \bibnamefont {Tan}},
  \bibinfo {author} {\bibfnamefont {S.}~\bibnamefont {Masuda}}, \bibinfo
  {author} {\bibfnamefont {J.}~\bibnamefont {Govenius}}, \bibinfo {author}
  {\bibfnamefont {R.~E.}\ \bibnamefont {Lake}}, \bibinfo {author}
  {\bibfnamefont {M.}~\bibnamefont {Jenei}}, \bibinfo {author} {\bibfnamefont
  {L.}~\bibnamefont {Gr{\"o}nberg}}, \bibinfo {author} {\bibfnamefont
  {J.}~\bibnamefont {Hassel}}, \bibinfo {author} {\bibfnamefont
  {S.}~\bibnamefont {Simbierowicz}}, \bibinfo {author} {\bibfnamefont
  {V.}~\bibnamefont {Vesterinen}}, \bibinfo {author} {\bibfnamefont
  {J.}~\bibnamefont {Tuorila}}, \bibinfo {author} {\bibfnamefont
  {T.}~\bibnamefont {Ala-Nissila}}, \ and\ \bibinfo {author} {\bibfnamefont
  {M.}~\bibnamefont {M{\"o}tt{\"o}nen}},\ }\bibfield  {title} {\enquote
  {\bibinfo {title} {Flux-tunable heat sink for quantum electric circuits},}\
  }\href {\doibase 10.1038/s41598-018-24449-1} {\bibfield  {journal} {\bibinfo
  {journal} {Scientific Reports}\ }\textbf {\bibinfo {volume} {8}},\ \bibinfo
  {pages} {6325} (\bibinfo {year} {2018})}\BibitemShut {NoStop}%
\bibitem [{\citenamefont {Senior}\ \emph {et~al.}(2020)\citenamefont {Senior},
  \citenamefont {Gubaydullin}, \citenamefont {Karimi}, \citenamefont
  {Peltonen}, \citenamefont {Ankerhold},\ and\ \citenamefont
  {Pekola}}]{Senior2020}%
  \BibitemOpen
  \bibfield  {author} {\bibinfo {author} {\bibfnamefont {J.}~\bibnamefont
  {Senior}}, \bibinfo {author} {\bibfnamefont {A.}~\bibnamefont {Gubaydullin}},
  \bibinfo {author} {\bibfnamefont {B.}~\bibnamefont {Karimi}}, \bibinfo
  {author} {\bibfnamefont {J.~T.}\ \bibnamefont {Peltonen}}, \bibinfo {author}
  {\bibfnamefont {J.}~\bibnamefont {Ankerhold}}, \ and\ \bibinfo {author}
  {\bibfnamefont {J.~P.}\ \bibnamefont {Pekola}},\ }\bibfield  {title}
  {\enquote {\bibinfo {title} {Heat rectification via a superconducting
  artificial atom},}\ }\href {\doibase 10.1038/s42005-020-0307-5} {\bibfield
  {journal} {\bibinfo  {journal} {Comm. Phys.}\ }\textbf {\bibinfo {volume}
  {3}},\ \bibinfo {pages} {40} (\bibinfo {year} {2020})}\BibitemShut {NoStop}%
\bibitem [{\citenamefont {Li}\ \emph {et~al.}(2006)\citenamefont {Li},
  \citenamefont {Wang},\ and\ \citenamefont {Casati}}]{li2006}%
  \BibitemOpen
  \bibfield  {author} {\bibinfo {author} {\bibfnamefont {B.}~\bibnamefont
  {Li}}, \bibinfo {author} {\bibfnamefont {L.}~\bibnamefont {Wang}}, \ and\
  \bibinfo {author} {\bibfnamefont {G.}~\bibnamefont {Casati}},\ }\bibfield
  {title} {\enquote {\bibinfo {title} {{Negative differential thermal
  resistance and thermal transistor}},}\ }\href {\doibase 10.1063/1.2191730}
  {\bibfield  {journal} {\bibinfo  {journal} {Applied Physics Letters}\
  }\textbf {\bibinfo {volume} {88}},\ \bibinfo {pages} {143501} (\bibinfo
  {year} {2006})}\BibitemShut {NoStop}%
\bibitem [{\citenamefont {Giazotto}\ \emph {et~al.}(2014)\citenamefont
  {Giazotto}, \citenamefont {Robinson}, \citenamefont {Moodera},\ and\
  \citenamefont {Bergeret}}]{giazotto2014}%
  \BibitemOpen
  \bibfield  {author} {\bibinfo {author} {\bibfnamefont {F.}~\bibnamefont
  {Giazotto}}, \bibinfo {author} {\bibfnamefont {J.~W.~A.}\ \bibnamefont
  {Robinson}}, \bibinfo {author} {\bibfnamefont {J.~S.}\ \bibnamefont
  {Moodera}}, \ and\ \bibinfo {author} {\bibfnamefont {F.~S.}\ \bibnamefont
  {Bergeret}},\ }\bibfield  {title} {\enquote {\bibinfo {title} {{Proposal for
  a phase-coherent thermoelectric transistor}},}\ }\href {\doibase
  10.1063/1.4893443} {\bibfield  {journal} {\bibinfo  {journal} {Applied
  Physics Letters}\ }\textbf {\bibinfo {volume} {105}},\ \bibinfo {pages}
  {062602} (\bibinfo {year} {2014})}\BibitemShut {NoStop}%
\bibitem [{\citenamefont {S\'anchez}\ \emph {et~al.}(2017)\citenamefont
  {S\'anchez}, \citenamefont {Thierschmann},\ and\ \citenamefont
  {Molenkamp}}]{sanchez2017}%
  \BibitemOpen
  \bibfield  {author} {\bibinfo {author} {\bibfnamefont {R.}~\bibnamefont
  {S\'anchez}}, \bibinfo {author} {\bibfnamefont {H.}~\bibnamefont
  {Thierschmann}}, \ and\ \bibinfo {author} {\bibfnamefont {L.~W.}\
  \bibnamefont {Molenkamp}},\ }\bibfield  {title} {\enquote {\bibinfo {title}
  {{All-thermal transistor based on stochastic switching}},}\ }\href {\doibase
  10.1103/PhysRevB.95.241401} {\bibfield  {journal} {\bibinfo  {journal} {Phys.
  Rev. B}\ }\textbf {\bibinfo {volume} {95}},\ \bibinfo {pages} {241401}
  (\bibinfo {year} {2017})}\BibitemShut {NoStop}%
\bibitem [{\citenamefont {Yang}\ \emph {et~al.}(2019)\citenamefont {Yang},
  \citenamefont {Elouard}, \citenamefont {Splettstoesser}, \citenamefont
  {Sothmann}, \citenamefont {S\'anchez},\ and\ \citenamefont
  {Jordan}}]{yang2019}%
  \BibitemOpen
  \bibfield  {author} {\bibinfo {author} {\bibfnamefont {J.}~\bibnamefont
  {Yang}}, \bibinfo {author} {\bibfnamefont {C.}~\bibnamefont {Elouard}},
  \bibinfo {author} {\bibfnamefont {J.}~\bibnamefont {Splettstoesser}},
  \bibinfo {author} {\bibfnamefont {B.}~\bibnamefont {Sothmann}}, \bibinfo
  {author} {\bibfnamefont {R.}~\bibnamefont {S\'anchez}}, \ and\ \bibinfo
  {author} {\bibfnamefont {A.~N.}\ \bibnamefont {Jordan}},\ }\bibfield  {title}
  {\enquote {\bibinfo {title} {{Thermal transistor and thermometer based on
  Coulomb-coupled conductors}},}\ }\href {\doibase 10.1103/PhysRevB.100.045418}
  {\bibfield  {journal} {\bibinfo  {journal} {Phys. Rev. B}\ }\textbf {\bibinfo
  {volume} {100}},\ \bibinfo {pages} {045418} (\bibinfo {year}
  {2019})}\BibitemShut {NoStop}%
\bibitem [{\citenamefont {Strambini}\ \emph {et~al.}(2014)\citenamefont
  {Strambini}, \citenamefont {Bergeret},\ and\ \citenamefont
  {Giazotto}}]{strambini2014}%
  \BibitemOpen
  \bibfield  {author} {\bibinfo {author} {\bibfnamefont {E.}~\bibnamefont
  {Strambini}}, \bibinfo {author} {\bibfnamefont {F.~S.}\ \bibnamefont
  {Bergeret}}, \ and\ \bibinfo {author} {\bibfnamefont {F.}~\bibnamefont
  {Giazotto}},\ }\bibfield  {title} {\enquote {\bibinfo {title} {{Proximity
  nanovalve with large phase-tunable thermal conductance}},}\ }\href {\doibase
  10.1063/1.4893759} {\bibfield  {journal} {\bibinfo  {journal} {Applied
  Physics Letters}\ }\textbf {\bibinfo {volume} {105}},\ \bibinfo {pages}
  {082601} (\bibinfo {year} {2014})}\BibitemShut {NoStop}%
\bibitem [{\citenamefont {Ronzani}\ \emph {et~al.}(2018)\citenamefont
  {Ronzani}, \citenamefont {Karimi}, \citenamefont {Senior}, \citenamefont
  {Chang}, \citenamefont {Peltonen}, \citenamefont {Chen},\ and\ \citenamefont
  {Pekola}}]{Ronzani2018}%
  \BibitemOpen
  \bibfield  {author} {\bibinfo {author} {\bibfnamefont {A.}~\bibnamefont
  {Ronzani}}, \bibinfo {author} {\bibfnamefont {B.}~\bibnamefont {Karimi}},
  \bibinfo {author} {\bibfnamefont {J.}~\bibnamefont {Senior}}, \bibinfo
  {author} {\bibfnamefont {Y.-C.}\ \bibnamefont {Chang}}, \bibinfo {author}
  {\bibfnamefont {J.~T.}\ \bibnamefont {Peltonen}}, \bibinfo {author}
  {\bibfnamefont {C.}~\bibnamefont {Chen}}, \ and\ \bibinfo {author}
  {\bibfnamefont {J.~P.}\ \bibnamefont {Pekola}},\ }\bibfield  {title}
  {\enquote {\bibinfo {title} {{Tunable photonic heat transport in a quantum
  heat valve}},}\ }\href {\doibase 10.1038/s41567-018-0199-4} {\bibfield
  {journal} {\bibinfo  {journal} {Nature Physics}\ }\textbf {\bibinfo {volume}
  {14}},\ \bibinfo {pages} {991--995} (\bibinfo {year} {2018})}\BibitemShut
  {NoStop}%
\bibitem [{\citenamefont {Giazotto}\ and\ \citenamefont
  {Mart{\ifmmode\acute{\imath}\else\'{\i}\fi}nez-P{\ifmmode\acute{e}\else\'{e}\fi}rez}(2012)}]{Giazotto2012Dec}%
  \BibitemOpen
  \bibfield  {author} {\bibinfo {author} {\bibfnamefont {F.}~\bibnamefont
  {Giazotto}}\ and\ \bibinfo {author} {\bibfnamefont {M.~J.}\ \bibnamefont
  {Mart{\ifmmode\acute{\imath}\else\'{\i}\fi}nez-P{\ifmmode\acute{e}\else\'{e}\fi}rez}},\
  }\bibfield  {title} {\enquote {\bibinfo {title} {{The Josephson heat
  interferometer}},}\ }\href {\doibase 10.1038/nature11702} {\bibfield
  {journal} {\bibinfo  {journal} {Nature}\ }\textbf {\bibinfo {volume} {492}},\
  \bibinfo {pages} {401--405} (\bibinfo {year} {2012})}\BibitemShut {NoStop}%
\bibitem [{\citenamefont {Mart\'{i}nez-P\'{e}rez}\ and\ \citenamefont
  {Giazotto}(2013{\natexlab{b}})}]{martinez2013b}%
  \BibitemOpen
  \bibfield  {author} {\bibinfo {author} {\bibfnamefont {M.~J.}\ \bibnamefont
  {Mart\'{i}nez-P\'{e}rez}}\ and\ \bibinfo {author} {\bibfnamefont
  {F.}~\bibnamefont {Giazotto}},\ }\bibfield  {title} {\enquote {\bibinfo
  {title} {Fully balanced heat interferometer},}\ }\href {\doibase
  10.1063/1.4794412} {\bibfield  {journal} {\bibinfo  {journal} {Applied
  Physics Letters}\ }\textbf {\bibinfo {volume} {102}},\ \bibinfo {pages}
  {092602} (\bibinfo {year} {2013}{\natexlab{b}})}\BibitemShut {NoStop}%
\bibitem [{\citenamefont {Hwang}\ \emph {et~al.}(2018)\citenamefont {Hwang},
  \citenamefont {Giazotto},\ and\ \citenamefont {Sothmann}}]{hwang18}%
  \BibitemOpen
  \bibfield  {author} {\bibinfo {author} {\bibfnamefont {S.-Y.}\ \bibnamefont
  {Hwang}}, \bibinfo {author} {\bibfnamefont {F.}~\bibnamefont {Giazotto}}, \
  and\ \bibinfo {author} {\bibfnamefont {B.}~\bibnamefont {Sothmann}},\
  }\bibfield  {title} {\enquote {\bibinfo {title} {{Phase-Coherent Heat
  Circulator Based on Multiterminal Josephson Junctions}},}\ }\href {\doibase
  10.1103/PhysRevApplied.10.044062} {\bibfield  {journal} {\bibinfo  {journal}
  {Phys. Rev. Applied}\ }\textbf {\bibinfo {volume} {10}},\ \bibinfo {pages}
  {044062} (\bibinfo {year} {2018})}\BibitemShut {NoStop}%
\bibitem [{\citenamefont {Viola}\ and\ \citenamefont
  {DiVincenzo}(2014)}]{Viola2014May}%
  \BibitemOpen
  \bibfield  {author} {\bibinfo {author} {\bibfnamefont {G.}~\bibnamefont
  {Viola}}\ and\ \bibinfo {author} {\bibfnamefont {D.~P.}\ \bibnamefont
  {DiVincenzo}},\ }\bibfield  {title} {\enquote {\bibinfo {title} {{Hall Effect
  Gyrators and Circulators}},}\ }\href {\doibase 10.1103/PhysRevX.4.021019}
  {\bibfield  {journal} {\bibinfo  {journal} {Phys. Rev. X}\ }\textbf {\bibinfo
  {volume} {4}},\ \bibinfo {pages} {021019} (\bibinfo {year}
  {2014})}\BibitemShut {NoStop}%
\bibitem [{\citenamefont {Bosco}\ \emph {et~al.}(2017)\citenamefont {Bosco},
  \citenamefont {Haupt},\ and\ \citenamefont {DiVincenzo}}]{Bosco2017Feb}%
  \BibitemOpen
  \bibfield  {author} {\bibinfo {author} {\bibfnamefont {S.}~\bibnamefont
  {Bosco}}, \bibinfo {author} {\bibfnamefont {F.}~\bibnamefont {Haupt}}, \ and\
  \bibinfo {author} {\bibfnamefont {D.~P.}\ \bibnamefont {DiVincenzo}},\
  }\bibfield  {title} {\enquote {\bibinfo {title} {{Self-Impedance-Matched
  Hall-Effect Gyrators and Circulators}},}\ }\href {\doibase
  10.1103/PhysRevApplied.7.024030} {\bibfield  {journal} {\bibinfo  {journal}
  {Phys. Rev. Appl.}\ }\textbf {\bibinfo {volume} {7}},\ \bibinfo {pages}
  {024030} (\bibinfo {year} {2017})}\BibitemShut {NoStop}%
\bibitem [{\citenamefont {Mahoney}\ \emph {et~al.}(2017)\citenamefont
  {Mahoney}, \citenamefont {Colless}, \citenamefont {Pauka}, \citenamefont
  {Hornibrook}, \citenamefont {Watson}, \citenamefont {Gardner}, \citenamefont
  {Manfra}, \citenamefont {Doherty},\ and\ \citenamefont
  {Reilly}}]{mahoney2017}%
  \BibitemOpen
  \bibfield  {author} {\bibinfo {author} {\bibfnamefont {A.~C.}\ \bibnamefont
  {Mahoney}}, \bibinfo {author} {\bibfnamefont {J.~I.}\ \bibnamefont
  {Colless}}, \bibinfo {author} {\bibfnamefont {S.~J.}\ \bibnamefont {Pauka}},
  \bibinfo {author} {\bibfnamefont {J.~M.}\ \bibnamefont {Hornibrook}},
  \bibinfo {author} {\bibfnamefont {J.~D.}\ \bibnamefont {Watson}}, \bibinfo
  {author} {\bibfnamefont {G.~C.}\ \bibnamefont {Gardner}}, \bibinfo {author}
  {\bibfnamefont {M.~J.}\ \bibnamefont {Manfra}}, \bibinfo {author}
  {\bibfnamefont {A.~C.}\ \bibnamefont {Doherty}}, \ and\ \bibinfo {author}
  {\bibfnamefont {D.~J.}\ \bibnamefont {Reilly}},\ }\bibfield  {title}
  {\enquote {\bibinfo {title} {{On-Chip Microwave Quantum Hall Circulator}},}\
  }\href {\doibase 10.1103/PhysRevX.7.011007} {\bibfield  {journal} {\bibinfo
  {journal} {Phys. Rev. X}\ }\textbf {\bibinfo {volume} {7}},\ \bibinfo {pages}
  {011007} (\bibinfo {year} {2017})}\BibitemShut {NoStop}%
\bibitem [{\citenamefont {Chapman}\ \emph {et~al.}(2017)\citenamefont
  {Chapman}, \citenamefont {Rosenthal}, \citenamefont {Kerckhoff},
  \citenamefont {Moores}, \citenamefont {Vale}, \citenamefont {Mates},
  \citenamefont {Hilton}, \citenamefont {Lalumi\`ere}, \citenamefont {Blais},\
  and\ \citenamefont {Lehnert}}]{chapman2017}%
  \BibitemOpen
  \bibfield  {author} {\bibinfo {author} {\bibfnamefont {B.~J.}\ \bibnamefont
  {Chapman}}, \bibinfo {author} {\bibfnamefont {E.~I.}\ \bibnamefont
  {Rosenthal}}, \bibinfo {author} {\bibfnamefont {J.}~\bibnamefont
  {Kerckhoff}}, \bibinfo {author} {\bibfnamefont {B.~A.}\ \bibnamefont
  {Moores}}, \bibinfo {author} {\bibfnamefont {L.~R.}\ \bibnamefont {Vale}},
  \bibinfo {author} {\bibfnamefont {J.~A.~B.}\ \bibnamefont {Mates}}, \bibinfo
  {author} {\bibfnamefont {G.~C.}\ \bibnamefont {Hilton}}, \bibinfo {author}
  {\bibfnamefont {K.}~\bibnamefont {Lalumi\`ere}}, \bibinfo {author}
  {\bibfnamefont {A.}~\bibnamefont {Blais}}, \ and\ \bibinfo {author}
  {\bibfnamefont {K.~W.}\ \bibnamefont {Lehnert}},\ }\bibfield  {title}
  {\enquote {\bibinfo {title} {{Widely Tunable On-Chip Microwave Circulator for
  Superconducting Quantum Circuits}},}\ }\href {\doibase
  10.1103/PhysRevX.7.041043} {\bibfield  {journal} {\bibinfo  {journal} {Phys.
  Rev. X}\ }\textbf {\bibinfo {volume} {7}},\ \bibinfo {pages} {041043}
  (\bibinfo {year} {2017})}\BibitemShut {NoStop}%
\bibitem [{\citenamefont {Wu}\ and\ \citenamefont
  {Ramamurthy}(2002)}]{Wu2002Jan}%
  \BibitemOpen
  \bibfield  {author} {\bibinfo {author} {\bibfnamefont {C.~H.}\ \bibnamefont
  {Wu}}\ and\ \bibinfo {author} {\bibfnamefont {D.}~\bibnamefont
  {Ramamurthy}},\ }\bibfield  {title} {\enquote {\bibinfo {title} {{Logic
  functions from three-terminal quantum resistor networks for electron wave
  computing}},}\ }\href {\doibase 10.1103/PhysRevB.65.075313} {\bibfield
  {journal} {\bibinfo  {journal} {Phys. Rev. B}\ }\textbf {\bibinfo {volume}
  {65}},\ \bibinfo {pages} {075313} (\bibinfo {year} {2002})}\BibitemShut
  {NoStop}%
\bibitem [{\citenamefont {Strambini}\ \emph {et~al.}(2009)\citenamefont
  {Strambini}, \citenamefont {Piazza}, \citenamefont {Biasiol}, \citenamefont
  {Sorba},\ and\ \citenamefont {Beltram}}]{Strambini2009May}%
  \BibitemOpen
  \bibfield  {author} {\bibinfo {author} {\bibfnamefont {E.}~\bibnamefont
  {Strambini}}, \bibinfo {author} {\bibfnamefont {V.}~\bibnamefont {Piazza}},
  \bibinfo {author} {\bibfnamefont {G.}~\bibnamefont {Biasiol}}, \bibinfo
  {author} {\bibfnamefont {L.}~\bibnamefont {Sorba}}, \ and\ \bibinfo {author}
  {\bibfnamefont {F.}~\bibnamefont {Beltram}},\ }\bibfield  {title} {\enquote
  {\bibinfo {title} {{Impact of classical forces and decoherence in
  multiterminal Aharonov-Bohm networks}},}\ }\href {\doibase
  10.1103/PhysRevB.79.195443} {\bibfield  {journal} {\bibinfo  {journal} {Phys.
  Rev. B}\ }\textbf {\bibinfo {volume} {79}},\ \bibinfo {pages} {195443}
  (\bibinfo {year} {2009})}\BibitemShut {NoStop}%
\bibitem [{\citenamefont {Maki}\ and\ \citenamefont
  {Griffin}(1965)}]{Maki1965Dec}%
  \BibitemOpen
  \bibfield  {author} {\bibinfo {author} {\bibfnamefont {K.}~\bibnamefont
  {Maki}}\ and\ \bibinfo {author} {\bibfnamefont {A.}~\bibnamefont {Griffin}},\
  }\bibfield  {title} {\enquote {\bibinfo {title} {{Entropy Transport Between
  Two Superconductors by Electron Tunneling}},}\ }\href {\doibase
  10.1103/PhysRevLett.15.921} {\bibfield  {journal} {\bibinfo  {journal} {Phys.
  Rev. Lett.}\ }\textbf {\bibinfo {volume} {15}},\ \bibinfo {pages} {921--923}
  (\bibinfo {year} {1965})}\BibitemShut {NoStop}%
\bibitem [{\citenamefont {Zhao}\ \emph {et~al.}(2003)\citenamefont {Zhao},
  \citenamefont {L{\ifmmode\ddot{o}\else\"{o}\fi}fwander},\ and\ \citenamefont
  {Sauls}}]{Zhao2003Aug}%
  \BibitemOpen
  \bibfield  {author} {\bibinfo {author} {\bibfnamefont {E.}~\bibnamefont
  {Zhao}}, \bibinfo {author} {\bibfnamefont {T.}~\bibnamefont
  {L{\ifmmode\ddot{o}\else\"{o}\fi}fwander}}, \ and\ \bibinfo {author}
  {\bibfnamefont {J.~A.}\ \bibnamefont {Sauls}},\ }\bibfield  {title} {\enquote
  {\bibinfo {title} {{Phase Modulated Thermal Conductance of Josephson Weak
  Links}},}\ }\href {\doibase 10.1103/PhysRevLett.91.077003} {\bibfield
  {journal} {\bibinfo  {journal} {Phys. Rev. Lett.}\ }\textbf {\bibinfo
  {volume} {91}},\ \bibinfo {pages} {077003} (\bibinfo {year}
  {2003})}\BibitemShut {NoStop}%
\bibitem [{\citenamefont {Virtanen}\ and\ \citenamefont
  {Giazotto}(2015)}]{Virtanen2015Feb}%
  \BibitemOpen
  \bibfield  {author} {\bibinfo {author} {\bibfnamefont {P.}~\bibnamefont
  {Virtanen}}\ and\ \bibinfo {author} {\bibfnamefont {F.}~\bibnamefont
  {Giazotto}},\ }\bibfield  {title} {\enquote {\bibinfo {title} {{Fluctuation
  of heat current in Josephson junctions}},}\ }\href {\doibase
  10.1063/1.4914077} {\bibfield  {journal} {\bibinfo  {journal} {AIP Adv.}\
  }\textbf {\bibinfo {volume} {5}},\ \bibinfo {pages} {027140} (\bibinfo {year}
  {2015})}\BibitemShut {NoStop}%
\bibitem [{\citenamefont {Hajiloo}\ \emph {et~al.}(2019)\citenamefont
  {Hajiloo}, \citenamefont {Hassler},\ and\ \citenamefont
  {Splettstoesser}}]{Hajiloo2019Jun}%
  \BibitemOpen
  \bibfield  {author} {\bibinfo {author} {\bibfnamefont {F.}~\bibnamefont
  {Hajiloo}}, \bibinfo {author} {\bibfnamefont {F.}~\bibnamefont {Hassler}}, \
  and\ \bibinfo {author} {\bibfnamefont {J.}~\bibnamefont {Splettstoesser}},\
  }\bibfield  {title} {\enquote {\bibinfo {title} {{Mesoscopic effects in the
  heat conductance of superconducting-normal-superconducting and
  normal-superconducting junctions}},}\ }\href {\doibase
  10.1103/PhysRevB.99.235422} {\bibfield  {journal} {\bibinfo  {journal} {Phys.
  Rev. B}\ }\textbf {\bibinfo {volume} {99}},\ \bibinfo {pages} {235422}
  (\bibinfo {year} {2019})}\BibitemShut {NoStop}%
\bibitem [{\citenamefont {Alhassid}(2000)}]{alhassid2000}%
  \BibitemOpen
  \bibfield  {author} {\bibinfo {author} {\bibfnamefont {Y.}~\bibnamefont
  {Alhassid}},\ }\bibfield  {title} {\enquote {\bibinfo {title} {{The
  statistical theory of quantum dots}},}\ }\href {\doibase
  10.1103/RevModPhys.72.895} {\bibfield  {journal} {\bibinfo  {journal} {Rev.
  Mod. Phys.}\ }\textbf {\bibinfo {volume} {72}},\ \bibinfo {pages} {895--968}
  (\bibinfo {year} {2000})}\BibitemShut {NoStop}%
\bibitem [{\citenamefont {Beenakker}(1997)}]{beenakker-review}%
  \BibitemOpen
  \bibfield  {author} {\bibinfo {author} {\bibfnamefont {C.~W.~J.}\
  \bibnamefont {Beenakker}},\ }\bibfield  {title} {\enquote {\bibinfo {title}
  {Random-matrix theory of quantum transport},}\ }\href {\doibase
  10.1103/RevModPhys.69.731} {\bibfield  {journal} {\bibinfo  {journal} {Rev.
  Mod. Phys.}\ }\textbf {\bibinfo {volume} {69}},\ \bibinfo {pages} {731--808}
  (\bibinfo {year} {1997})}\BibitemShut {NoStop}%
\bibitem [{\citenamefont {Blanter}\ and\ \citenamefont
  {B{\ifmmode\ddot{u}\else\"{u}\fi}ttiker}(2000)}]{Blanter2000}%
  \BibitemOpen
  \bibfield  {author} {\bibinfo {author} {\bibfnamefont {{\relax Ya}.~M.}\
  \bibnamefont {Blanter}}\ and\ \bibinfo {author} {\bibfnamefont
  {M.}~\bibnamefont {B{\ifmmode\ddot{u}\else\"{u}\fi}ttiker}},\ }\bibfield
  {title} {\enquote {\bibinfo {title} {{Shot noise in mesoscopic
  conductors}},}\ }\href {\doibase 10.1016/S0370-1573(99)00123-4} {\bibfield
  {journal} {\bibinfo  {journal} {Phys. Rep.}\ }\textbf {\bibinfo {volume}
  {336}},\ \bibinfo {pages} {1--166} (\bibinfo {year} {2000})}\BibitemShut
  {NoStop}%
\bibitem [{\citenamefont {Nazarov}\ and\ \citenamefont
  {Blanter}(2009)}]{nazarov_book}%
  \BibitemOpen
  \bibfield  {author} {\bibinfo {author} {\bibfnamefont {Y.~V.}\ \bibnamefont
  {Nazarov}}\ and\ \bibinfo {author} {\bibfnamefont {Y.~M.}\ \bibnamefont
  {Blanter}},\ }\href {\doibase 10.1017/CBO9780511626906} {\emph {\bibinfo
  {title} {Quantum Transport: Introduction to Nanoscience}}}\ (\bibinfo
  {publisher} {Cambridge University Press},\ \bibinfo {year}
  {2009})\BibitemShut {NoStop}%
\bibitem [{\citenamefont {Spilla}\ \emph {et~al.}(2014)\citenamefont {Spilla},
  \citenamefont {Hassler},\ and\ \citenamefont
  {Splettstoesser}}]{Spilla2014Apr}%
  \BibitemOpen
  \bibfield  {author} {\bibinfo {author} {\bibfnamefont {S.}~\bibnamefont
  {Spilla}}, \bibinfo {author} {\bibfnamefont {F.}~\bibnamefont {Hassler}}, \
  and\ \bibinfo {author} {\bibfnamefont {J.}~\bibnamefont {Splettstoesser}},\
  }\bibfield  {title} {\enquote {\bibinfo {title} {{Measurement and dephasing
  of a flux qubit due to heat currents}},}\ }\href {\doibase
  10.1088/1367-2630/16/4/045020} {\bibfield  {journal} {\bibinfo  {journal}
  {New J. Phys.}\ }\textbf {\bibinfo {volume} {16}},\ \bibinfo {pages} {045020}
  (\bibinfo {year} {2014})}\BibitemShut {NoStop}%
\bibitem [{\citenamefont {Spilla}\ \emph {et~al.}(2015)\citenamefont {Spilla},
  \citenamefont {Hassler}, \citenamefont {Napoli},\ and\ \citenamefont
  {Splettstoesser}}]{Spilla2015Jun}%
  \BibitemOpen
  \bibfield  {author} {\bibinfo {author} {\bibfnamefont {S.}~\bibnamefont
  {Spilla}}, \bibinfo {author} {\bibfnamefont {F.}~\bibnamefont {Hassler}},
  \bibinfo {author} {\bibfnamefont {A.}~\bibnamefont {Napoli}}, \ and\ \bibinfo
  {author} {\bibfnamefont {J.}~\bibnamefont {Splettstoesser}},\ }\bibfield
  {title} {\enquote {\bibinfo {title} {{Dephasing due to quasiparticle
  tunneling in fluxonium qubits: a phenomenological approach}},}\ }\href
  {\doibase 10.1088/1367-2630/17/6/065012} {\bibfield  {journal} {\bibinfo
  {journal} {New J. Phys.}\ }\textbf {\bibinfo {volume} {17}},\ \bibinfo
  {pages} {065012} (\bibinfo {year} {2015})}\BibitemShut {NoStop}%
\bibitem [{\citenamefont {Meyer}\ and\ \citenamefont {Houzet}(2017)}]{meyer17}%
  \BibitemOpen
  \bibfield  {author} {\bibinfo {author} {\bibfnamefont {J.~S.}\ \bibnamefont
  {Meyer}}\ and\ \bibinfo {author} {\bibfnamefont {M.}~\bibnamefont {Houzet}},\
  }\bibfield  {title} {\enquote {\bibinfo {title} {{Nontrivial Chern Numbers in
  Three-Terminal Josephson Junctions}},}\ }\href {\doibase
  10.1103/PhysRevLett.119.136807} {\bibfield  {journal} {\bibinfo  {journal}
  {Phys. Rev. Lett.}\ }\textbf {\bibinfo {volume} {119}},\ \bibinfo {pages}
  {136807} (\bibinfo {year} {2017})}\BibitemShut {NoStop}%
\bibitem [{\citenamefont {Beenakker}(1991)}]{Beenakker1991Dec}%
  \BibitemOpen
  \bibfield  {author} {\bibinfo {author} {\bibfnamefont {C.~W.~J.}\
  \bibnamefont {Beenakker}},\ }\bibfield  {title} {\enquote {\bibinfo {title}
  {{Universal limit of critical-current fluctuations in mesoscopic Josephson
  junctions}},}\ }\href {\doibase 10.1103/PhysRevLett.67.3836} {\bibfield
  {journal} {\bibinfo  {journal} {Phys. Rev. Lett.}\ }\textbf {\bibinfo
  {volume} {67}},\ \bibinfo {pages} {3836--3839} (\bibinfo {year}
  {1991})}\BibitemShut {NoStop}%
\bibitem [{\citenamefont {Blonder}\ \emph {et~al.}(1982)\citenamefont
  {Blonder}, \citenamefont {Tinkham},\ and\ \citenamefont
  {Klapwijk}}]{blonder82}%
  \BibitemOpen
  \bibfield  {author} {\bibinfo {author} {\bibfnamefont {G.~E.}\ \bibnamefont
  {Blonder}}, \bibinfo {author} {\bibfnamefont {M.}~\bibnamefont {Tinkham}}, \
  and\ \bibinfo {author} {\bibfnamefont {T.~M.}\ \bibnamefont {Klapwijk}},\
  }\bibfield  {title} {\enquote {\bibinfo {title} {{Transition from metallic to
  tunneling regimes in superconducting microconstrictions: Excess current,
  charge imbalance, and supercurrent conversion}},}\ }\href {\doibase
  10.1103/PhysRevB.25.4515} {\bibfield  {journal} {\bibinfo  {journal} {Phys.
  Rev. B}\ }\textbf {\bibinfo {volume} {25}},\ \bibinfo {pages} {4515--4532}
  (\bibinfo {year} {1982})}\BibitemShut {NoStop}%
\bibitem [{\citenamefont {Moskalets}(2011)}]{moskalets-book}%
  \BibitemOpen
  \bibfield  {author} {\bibinfo {author} {\bibfnamefont {M.~V.}\ \bibnamefont
  {Moskalets}},\ }\href {\doibase 10.1142/p822} {\emph {\bibinfo {title}
  {{Scattering Matrix Approach to Non-Stationary Quantum Transport}}}}\
  (\bibinfo  {publisher} {Imperial College Press},\ \bibinfo {year}
  {2011})\BibitemShut {NoStop}%
\bibitem [{\citenamefont {Brouwer}\ and\ \citenamefont
  {Beenakker}(1996{\natexlab{a}})}]{brouwer96}%
  \BibitemOpen
  \bibfield  {author} {\bibinfo {author} {\bibfnamefont {P.~W.}\ \bibnamefont
  {Brouwer}}\ and\ \bibinfo {author} {\bibfnamefont {C.~W.~J.}\ \bibnamefont
  {Beenakker}},\ }\bibfield  {title} {\enquote {\bibinfo {title}
  {{Phase-dependent magnetoconductance fluctuations in a chaotic Josephson
  junction}},}\ }\href {\doibase 10.1103/PhysRevB.54.R12705} {\bibfield
  {journal} {\bibinfo  {journal} {Phys. Rev. B}\ }\textbf {\bibinfo {volume}
  {54}},\ \bibinfo {pages} {R12705--R12708} (\bibinfo {year}
  {1996}{\natexlab{a}})}\BibitemShut {NoStop}%
\bibitem [{\citenamefont {Brouwer}\ and\ \citenamefont
  {Beenakker}(1996{\natexlab{b}})}]{brouwer96b}%
  \BibitemOpen
  \bibfield  {author} {\bibinfo {author} {\bibfnamefont {P.~W.}\ \bibnamefont
  {Brouwer}}\ and\ \bibinfo {author} {\bibfnamefont {C.~W.~J.}\ \bibnamefont
  {Beenakker}},\ }\bibfield  {title} {\enquote {\bibinfo {title} {Diagrammatic
  method of integration over the unitary group, with applications to quantum
  transport in mesoscopic systems},}\ }\href {\doibase 10.1063/1.531667}
  {\bibfield  {journal} {\bibinfo  {journal} {Journal of Mathematical Physics}\
  }\textbf {\bibinfo {volume} {37}},\ \bibinfo {pages} {4904--4934} (\bibinfo
  {year} {1996}{\natexlab{b}})}\BibitemShut {NoStop}%
\bibitem [{\citenamefont {Bronzan}(1988)}]{bronzan1988}%
  \BibitemOpen
  \bibfield  {author} {\bibinfo {author} {\bibfnamefont {J.~B.}\ \bibnamefont
  {Bronzan}},\ }\bibfield  {title} {\enquote {\bibinfo {title}
  {{Parametrization of SU(3)}},}\ }\href {\doibase 10.1103/PhysRevD.38.1994}
  {\bibfield  {journal} {\bibinfo  {journal} {Phys. Rev. D}\ }\textbf {\bibinfo
  {volume} {38}},\ \bibinfo {pages} {1994} (\bibinfo {year}
  {1988})}\BibitemShut {NoStop}%
\bibitem [{\citenamefont {Krive}\ \emph {et~al.}(2001)\citenamefont {Krive},
  \citenamefont {Bogachek}, \citenamefont {Scherbakov},\ and\ \citenamefont
  {Landman}}]{Krive2001Nov}%
  \BibitemOpen
  \bibfield  {author} {\bibinfo {author} {\bibfnamefont {I.~V.}\ \bibnamefont
  {Krive}}, \bibinfo {author} {\bibfnamefont {E.~N.}\ \bibnamefont {Bogachek}},
  \bibinfo {author} {\bibfnamefont {A.~G.}\ \bibnamefont {Scherbakov}}, \ and\
  \bibinfo {author} {\bibfnamefont {U.}~\bibnamefont {Landman}},\ }\bibfield
  {title} {\enquote {\bibinfo {title} {{Heat current fluctuations in quantum
  wires}},}\ }\href {\doibase 10.1103/PhysRevB.64.233304} {\bibfield  {journal}
  {\bibinfo  {journal} {Phys. Rev. B}\ }\textbf {\bibinfo {volume} {64}},\
  \bibinfo {pages} {233304} (\bibinfo {year} {2001})}\BibitemShut {NoStop}%
\bibitem [{\citenamefont {Saito}\ and\ \citenamefont
  {Dhar}(2007)}]{Saito2007Oct}%
  \BibitemOpen
  \bibfield  {author} {\bibinfo {author} {\bibfnamefont {K.}~\bibnamefont
  {Saito}}\ and\ \bibinfo {author} {\bibfnamefont {A.}~\bibnamefont {Dhar}},\
  }\bibfield  {title} {\enquote {\bibinfo {title} {{Fluctuation Theorem in
  Quantum Heat Conduction}},}\ }\href {\doibase 10.1103/PhysRevLett.99.180601}
  {\bibfield  {journal} {\bibinfo  {journal} {Phys. Rev. Lett.}\ }\textbf
  {\bibinfo {volume} {99}},\ \bibinfo {pages} {180601} (\bibinfo {year}
  {2007})}\BibitemShut {NoStop}%
\bibitem [{\citenamefont {Saito}\ and\ \citenamefont
  {Dhar}(2008)}]{Saito2008Jul}%
  \BibitemOpen
  \bibfield  {author} {\bibinfo {author} {\bibfnamefont {K.}~\bibnamefont
  {Saito}}\ and\ \bibinfo {author} {\bibfnamefont {A.}~\bibnamefont {Dhar}},\
  }\bibfield  {title} {\enquote {\bibinfo {title} {{Erratum: Fluctuation
  Theorem in Quantum Heat Conduction [Phys. Rev. Lett. 99, 180601 (2007)]}},}\
  }\href {\doibase 10.1103/PhysRevLett.101.049902} {\bibfield  {journal}
  {\bibinfo  {journal} {Phys. Rev. Lett.}\ }\textbf {\bibinfo {volume} {101}},\
  \bibinfo {pages} {049902} (\bibinfo {year} {2008})}\BibitemShut {NoStop}%
\bibitem [{\citenamefont {Cr{\ifmmode\acute{e}\else\'{e}\fi}pieux}\ and\
  \citenamefont {Michelini}(2014)}]{Crepieux2014Dec}%
  \BibitemOpen
  \bibfield  {author} {\bibinfo {author} {\bibfnamefont {A.}~\bibnamefont
  {Cr{\ifmmode\acute{e}\else\'{e}\fi}pieux}}\ and\ \bibinfo {author}
  {\bibfnamefont {F.}~\bibnamefont {Michelini}},\ }\bibfield  {title} {\enquote
  {\bibinfo {title} {{Mixed, charge and heat noises in thermoelectric
  nanosystems}},}\ }\href {\doibase 10.1088/0953-8984/27/1/015302} {\bibfield
  {journal} {\bibinfo  {journal} {J. Phys.: Condens. Matter}\ }\textbf
  {\bibinfo {volume} {27}},\ \bibinfo {pages} {015302} (\bibinfo {year}
  {2014})}\BibitemShut {NoStop}%
\bibitem [{\citenamefont
  {Cr{\ifmmode\acute{e}\else\'{e}\fi}pieux}(2020)}]{Crepieux2020Jul}%
  \BibitemOpen
  \bibfield  {author} {\bibinfo {author} {\bibfnamefont {A.}~\bibnamefont
  {Cr{\ifmmode\acute{e}\else\'{e}\fi}pieux}},\ }\bibfield  {title} {\enquote
  {\bibinfo {title} {{Electronic heat current fluctuations in a quantum
  dot}},}\ }\href {https://arxiv.org/abs/2007.14827v1} {\bibfield  {journal}
  {\bibinfo  {journal} {arXiv}\ } (\bibinfo {year} {2020})},\ \Eprint
  {http://arxiv.org/abs/2007.14827} {2007.14827} \BibitemShut {NoStop}%
\bibitem [{\citenamefont {Samuel}(1980)}]{stuart1980}%
  \BibitemOpen
  \bibfield  {author} {\bibinfo {author} {\bibfnamefont {S.}~\bibnamefont
  {Samuel}},\ }\bibfield  {title} {\enquote {\bibinfo {title} {{$U(N)$
  Integrals, $1/N$, and the De Wit–’t Hooft anomalies}},}\ }\href {\doibase
  10.1063/1.524386} {\bibfield  {journal} {\bibinfo  {journal} {Journal of
  Mathematical Physics}\ }\textbf {\bibinfo {volume} {21}},\ \bibinfo {pages}
  {2695--2703} (\bibinfo {year} {1980})}\BibitemShut {NoStop}%
\end{thebibliography}%

\end{document}